\documentclass[preprint2]{emulateapj}

\usepackage{amsmath}
\usepackage{natbib}
\usepackage{graphicx}
\usepackage{epsf}
\usepackage{subfigure}
\usepackage{color}
\usepackage{threeparttable}
\usepackage{comment}
\usepackage{epsfig}
\usepackage{xspace}
\usepackage{hyperref}
\usepackage[usenames,dvipsnames,svgnames]{xcolor}
\usepackage{latexsym}
\DeclareGraphicsExtensions{.jpg,.pdf,.png,.eps,.ps}

 \newcommand{\LCDM}{\mbox{$\Lambda$CDM}\xspace}

 \newcommand{\ltsima}{$\; \buildrel < \over \sim \;$}
 \newcommand{\ltsim}{\lower.5ex\hbox{\ltsima}}

 \newcommand{\sqdeg}{\ensuremath{\mathrm{deg}^2}}

 \newcommand{\alens}{\ensuremath{A_\mathrm{lens}}}
 \newcommand{\adust}{\ensuremath{A_\mathrm{dust}}}

 \newcommand{\cldust}{\ensuremath{C_\ell^\mathrm{dust}}}

 \newcommand{\muksq}{\ensuremath{\mu{\rm K}^2}}

 \newcommand{\planck}{\textit{Planck}}

 \newcommand{\degs}{\ensuremath{\mathrm{deg}^2}}
 
 \newcommand{\simleq}{{\raise.0ex\hbox{$\mathchar"013C$}\mkern-14mu \lower1.2ex\hbox{$\mathchar"0218$}}}
 \newcommand{\simgeq}{{\raise.0ex\hbox{$\mathchar"013E$}\mkern-14mu \lower1.2ex\hbox{$\mathchar"0218$}}}
\newcommand{\be}{\begin{equation}}
\newcommand{\ee}{\end{equation}}
\newcommand{\TtoB}{$T \rightarrow B$}
\newcommand{\EtoB}{$E \rightarrow B$}

\begin{document}
\title{Measurements of Sub-degree B-mode Polarization in the Cosmic Microwave Background from 100 Square Degrees of SPTpol Data}

\def\Stanford{1}
\def\KIPAC{2}
\def\KICPChicago{3}
\def\PhysicsUChicago{4}
\def\Berkeley{5}
\def\ColoradoAPS{6}
\def\Cardiff{7}
\def\UChicago{8}
\def\NIST{9}
\def\McGill{10}
\def\ArgonneHEP{11}
\def\AAUChicago{12}
\def\FNAL{13}
\def\EFIChicago{14}
\def\UKZN{15}
\def\SLAC{16}
\def\Caltech{17}
\def\CIFAR{18}
\def\ColoradoPhys{19}
\def\Davis{20}
\def\LBNL{21}
\def\Arizona{22}
\def\Michigan{23}
\def\CaseWestern{24}
\def\ArgonneMSD{25}
\def\Minnesota{26}
\def\Melbourne{27}
\def\ArtInstChicago{28}
\def\ThreeSpeedLogic{29}
\def\CfA{30}
\def\Dunlap{31}
\def\UToronto{32}
\def\illast{33}
\def\illphy{34}
\def\BCCP{35}


 \author{
  R.~Keisler\altaffilmark{\Stanford,\KIPAC},
  S.~Hoover\altaffilmark{\KICPChicago,\PhysicsUChicago},
  N.~Harrington\altaffilmark{\Berkeley},
  J.~W.~Henning\altaffilmark{\KICPChicago,\ColoradoAPS},
  P.~A.~R.~Ade\altaffilmark{\Cardiff},
  K.~A.~Aird\altaffilmark{\UChicago},
  J.~E.~Austermann\altaffilmark{\ColoradoAPS, \NIST},
  J.~A.~Beall\altaffilmark{\NIST},
  A.~N.~Bender\altaffilmark{\McGill,\ArgonneHEP, \KICPChicago},
  B.~A.~Benson\altaffilmark{\KICPChicago,\AAUChicago,\FNAL},
  L.~E.~Bleem\altaffilmark{\KICPChicago,\PhysicsUChicago,\ArgonneHEP},
  J.~E.~Carlstrom\altaffilmark{\KICPChicago,\PhysicsUChicago,\ArgonneHEP,\AAUChicago,\EFIChicago},
  C.~L.~Chang\altaffilmark{\KICPChicago,\ArgonneHEP,\AAUChicago},
  H.~C.~Chiang\altaffilmark{\UKZN},
  H-M.~Cho\altaffilmark{\SLAC},
  R.~Citron\altaffilmark{\KICPChicago},
  T.~M.~Crawford\altaffilmark{\KICPChicago,\AAUChicago},
  A.~T.~Crites\altaffilmark{\KICPChicago,\AAUChicago,\Caltech},
  T.~de~Haan\altaffilmark{\Berkeley}, 
  M.~A.~Dobbs\altaffilmark{\McGill,\CIFAR},
  W.~Everett\altaffilmark{\ColoradoAPS},
  J.~Gallicchio\altaffilmark{\KICPChicago},
  J.~Gao\altaffilmark{\NIST},
  E.~M.~George\altaffilmark{\Berkeley},
  A.~Gilbert\altaffilmark{\McGill},
  N.~W.~Halverson\altaffilmark{\ColoradoAPS,\ColoradoPhys},
  D.~Hanson\altaffilmark{\McGill},
  G.~C.~Hilton\altaffilmark{\NIST},
  G.~P.~Holder\altaffilmark{\McGill},
  W.~L.~Holzapfel\altaffilmark{\Berkeley},
  Z.~Hou\altaffilmark{\KICPChicago},
  J.~D.~Hrubes\altaffilmark{\UChicago},
  N.~Huang\altaffilmark{\Berkeley},
  J.~Hubmayr\altaffilmark{\NIST},
  K.~D.~Irwin\altaffilmark{\Stanford, \SLAC},
  L.~Knox\altaffilmark{\Davis},
  A.~T.~Lee\altaffilmark{\Berkeley,\LBNL},
  E.~M.~Leitch\altaffilmark{\KICPChicago,\AAUChicago},
  D.~Li\altaffilmark{\NIST,\SLAC},
  D.~Luong-Van\altaffilmark{\UChicago},
  D.~P.~Marrone\altaffilmark{\Arizona},
  J.~J.~McMahon\altaffilmark{\Michigan},
  J.~Mehl\altaffilmark{\KICPChicago,\ArgonneHEP},
  S.~S.~Meyer\altaffilmark{\KICPChicago,\PhysicsUChicago,\AAUChicago,\EFIChicago},
  L.~Mocanu\altaffilmark{\KICPChicago,\AAUChicago},
  T.~Natoli\altaffilmark{\KICPChicago,\PhysicsUChicago},
  J.~P.~Nibarger\altaffilmark{\NIST},
  V.~Novosad\altaffilmark{\ArgonneMSD},
  S.~Padin\altaffilmark{\Caltech},
  C.~Pryke\altaffilmark{\Minnesota},
  C.~L.~Reichardt\altaffilmark{\Berkeley,\Melbourne},
  J.~E.~Ruhl\altaffilmark{\CaseWestern},
  B.~R.~Saliwanchik\altaffilmark{\CaseWestern},
  J.T.~Sayre\altaffilmark{\CaseWestern},
  K.~K.~Schaffer\altaffilmark{\KICPChicago,\EFIChicago,\ArtInstChicago},
  E.~Shirokoff\altaffilmark{\KICPChicago, \AAUChicago},
  G.~Smecher\altaffilmark{\McGill,\ThreeSpeedLogic},
  A.~A.~Stark\altaffilmark{\CfA},
  K.~T.~Story\altaffilmark{\KICPChicago,\PhysicsUChicago},
  C.~Tucker\altaffilmark{\Cardiff},
  K.~Vanderlinde\altaffilmark{\Dunlap,\UToronto},
  J.~D.~Vieira\altaffilmark{\illast,\illphy},
  G.~Wang\altaffilmark{\ArgonneHEP},
  N.~Whitehorn\altaffilmark{\Berkeley},
  V.~Yefremenko\altaffilmark{\ArgonneHEP},
  and 
  O.~Zahn\altaffilmark{\BCCP}
 }

\altaffiltext{\Stanford}{Department of Physics, Stanford University, 382 Via Pueblo Mall, Stanford, CA 94305}
\altaffiltext{\KIPAC}{Kavli Institute for Particle Astrophysics and Cosmology, Stanford University, 452 Lomita Mall, Stanford, CA 94305}
\altaffiltext{\KICPChicago}{Kavli Institute for Cosmological Physics, University of Chicago, 5640 South Ellis Avenue, Chicago, IL, USA 60637}
\altaffiltext{\PhysicsUChicago}{Department of Physics, University of Chicago, 5640 South Ellis Avenue, Chicago, IL, USA 60637}
\altaffiltext{\Berkeley}{Department of Physics, University of California, Berkeley, CA, USA 94720}
\altaffiltext{\ColoradoAPS}{Department of Astrophysical and Planetary Sciences, University of Colorado, Boulder, CO, USA 80309}
\altaffiltext{\Cardiff}{School of Physics and Astronomy, Cardiff University, CF24 3AA, United Kingdom}
\altaffiltext{\UChicago}{University of Chicago, 5640 South Ellis Avenue, Chicago, IL, USA 60637}
\altaffiltext{\NIST}{NIST Quantum Devices Group, 325 Broadway Mailcode 817.03, Boulder, CO, USA 80305}
\altaffiltext{\McGill}{Department of Physics, McGill University, 3600 Rue University, Montreal, Quebec H3A 2T8, Canada}
\altaffiltext{\ArgonneHEP}{High Energy Physics Division, Argonne National Laboratory,9700 S. Cass Avenue, Argonne, IL, USA 60439}
\altaffiltext{\AAUChicago}{Department of Astronomy and Astrophysics, University of Chicago, 5640 South Ellis Avenue, Chicago, IL, USA 60637}
\altaffiltext{\FNAL}{Fermi National Accelerator Laboratory, MS209, P.O. Box 500, Batavia, IL 60510}
\altaffiltext{\EFIChicago}{Enrico Fermi Institute, University of Chicago, 5640 South Ellis Avenue, Chicago, IL, USA 60637}
\altaffiltext{\UKZN}{School of Mathematics, Statistics \& Computer Science, University of KwaZulu-Natal, Durban, South Africa}
\altaffiltext{\SLAC}{SLAC National Accelerator Laboratory, 2575 Sand Hill Road, Menlo Park, CA 94025}
\altaffiltext{\Caltech}{California Institute of Technology, MS 367-17, 1200 E. California Blvd., Pasadena, CA, USA 91125}
\altaffiltext{\CIFAR}{Canadian Institute for Advanced Research, CIFAR Program in Cosmology and Gravity, Toronto, ON, M5G 1Z8, Canada}
\altaffiltext{\ColoradoPhys}{Department of Physics, University of Colorado, Boulder, CO, USA 80309}
\altaffiltext{\Davis}{Department of Physics, University of California, One Shields Avenue, Davis, CA, USA 95616}
\altaffiltext{\LBNL}{Physics Division, Lawrence Berkeley National Laboratory, Berkeley, CA, USA 94720}
\altaffiltext{\Arizona}{Steward Observatory, University of Arizona, 933 North Cherry Avenue, Tucson, AZ 85721, USA}
\altaffiltext{\Michigan}{Department of Physics, University of Michigan, 450 Church Street, Ann  Arbor, MI, USA 48109}
\altaffiltext{\CaseWestern}{Physics Department, Center for Education and Research in Cosmology and Astrophysics, Case Western Reserve University,Cleveland, OH, USA 44106}
\altaffiltext{\ArgonneMSD}{Materials Sciences Division, Argonne National Laboratory,9700 S. Cass Avenue, Argonne, IL, USA 60439}
\altaffiltext{\Minnesota}{School of Physics and Astronomy, University of Minnesota, 116 Church Street S.E. Minneapolis, MN, USA 55455}
\altaffiltext{\Melbourne}{School of Physics, University of Melbourne, Parkville, VIC 3010, Australia}
\altaffiltext{\ArtInstChicago}{Liberal Arts Department, School of the Art Institute of Chicago, 112 S Michigan Ave, Chicago, IL, USA 60603}
\altaffiltext{\ThreeSpeedLogic}{Three-Speed Logic, Inc., Vancouver, B.C., V6A 2J8, Canada}
\altaffiltext{\CfA}{Harvard-Smithsonian Center for Astrophysics, 60 Garden Street, Cambridge, MA, USA 02138}
\altaffiltext{\Dunlap}{Dunlap Institute for Astronomy \& Astrophysics, University of Toronto, 50 St George St, Toronto, ON, M5S 3H4, Canada}
\altaffiltext{\UToronto}{Department of Astronomy \& Astrophysics, University of Toronto, 50 St George St, Toronto, ON, M5S 3H4, Canada}
\altaffiltext{\illast}{Astronomy Department, University of Illinois at Urbana-Champaign, 1002 W.\ Green Street, Urbana, IL 61801, USA}
\altaffiltext{\illphy}{Department of Physics, University of Illinois Urbana-Champaign, 1110 W.\ Green Street, Urbana, IL 61801, USA}
\altaffiltext{\BCCP}{Berkeley Center for Cosmological Physics, Department of Physics, University of California, and Lawrence Berkeley National Laboratory, Berkeley, CA, USA 94720}

\email{rkeisler@stanford.edu}

\begin{abstract}

We present a measurement of the $B$-mode polarization 
power spectrum (the $BB$ spectrum) from 100 \degs\ of sky
observed with SPTpol, a polarization-sensitive 
receiver currently installed on the South Pole Telescope.
The observations used in this work were taken during 2012 and early 2013
and include data in spectral bands centered at 95 and 150\,GHz. 
We report the $BB$ spectrum in five bins in multipole 
space, spanning the range $300 \le \ell \le 2300$, 
and for three spectral combinations:
$95\,\mathrm{GHz} \times 95\,\mathrm{GHz}$, 
$95\,\mathrm{GHz} \times 150\,\mathrm{GHz}$, and
$150\,\mathrm{GHz} \times 150\,\mathrm{GHz}$.
We subtract small ($< 0.5 \sigma$ in units of statistical uncertainty) biases from these spectra and 
account for the uncertainty in those biases.
The resulting power spectra are inconsistent with zero power 
but consistent with predictions for the $BB$ spectrum arising from the gravitational lensing of $E$-mode polarization.
If we assume no other source of $BB$ power besides lensed $B$ modes, 
we determine a preference for lensed $B$ modes
of $4.9 \sigma$.
After marginalizing over tensor power and foregrounds, namely polarized emission from galactic dust and extragalactic sources, 
this significance is $4.3 \sigma$.
Fitting for a single parameter, \alens, that multiplies the predicted lensed $B$-mode 
spectrum, and marginalizing over tensor power and foregrounds,
we find $\alens = 1.08 \pm 0.26$, indicating that our measured spectra are consistent
with the signal expected from gravitational lensing.
The data presented here provide the best measurement  to date of the $B$-mode power spectrum on these angular scales.
\end{abstract}


\maketitle

\section{Introduction}
\label{sec:intro}
\setcounter{footnote}{0}

Measurements of the cosmic microwave background (CMB), the oldest light in the universe, 
contain a wealth of cosmological information, informing our understanding of physical 
processes across nearly all of cosmic time
(see \citealt{hu02b} for a review). 
The majority of CMB photons last interacted electromagnetically with matter at the epoch of recombination
($z \sim 1000$), and it is that period of cosmic history that is most tightly constrained by 
CMB observations.
However, the interaction between CMB photons and matter at lower redshifts
encodes information about the more recent history of the universe.
In particular, the bending of CMB photon trajectories due to gravitational lensing enables 
reconstruction of the gravitational potential between recombination and $z=0$.
Furthermore, the imprint of gravitational waves on the polarization of the CMB  
has the potential to probe the absolute earliest moments of cosmic time.

Current CMB-derived constraints on cosmological parameters rely primarily on information
from the angular power spectrum of the CMB temperature fluctuations 
\citep[the $TT$ spectrum, e.g.,][]{planck13-16}.  However, the polarization of the CMB holds 
a wealth of potential information that is just beginning to be exploited. 
As with any headless vector field on the sphere, the linear
polarization of the CMB\footnote{The circular polarization of the CMB is 
observed to be extremely small, as expected.  See e.g. \citealt{mainini13}.}
can be decomposed into a curl-free component and a divergence-free component,
often called ``$E$ modes'' and ``$B$ modes,'' respectively, after the analogous fields in 
electrodynamics. The primary mechanism responsible for CMB polarization is
Thomson scattering between electrons and CMB photons with an anisotropic temperature
distribution \citep[e.g.,][]{hu97d}. Scalar density perturbations at the epoch of 
recombination produce only $E$-mode polarization to first order via this mechanism.
$B$-mode polarization in the CMB can be produced by vector perturbations (primordial vorticity) 
or tensor perturbations (gravitational waves) and by the distortion of $E$ modes through 
gravitational lensing by matter between the last scattering surface and the observer 
\citep{seljak97,kamionkowski97,zaldarriaga98}.

The search for $B$-mode polarization in the CMB has been a topic of particular interest
 because the most successful model for explaining many 
of the observed features of the universe, the paradigm of cosmic inflation,
predicts the existence of a background of gravitational waves \citep[e.g.,][]{abazajian15a}. 
These gravitational waves
leave their imprint on the CMB through scattering at the epoch of recombination (and again
at the epoch of reionization) through a contribution to the temperature, $E$-mode polarization,
and---most importantly---$B$-mode polarization of the CMB. The gravity-wave contribution to the temperature 
and $E$-mode power spectra is already constrained to be too small to measure (due to 
cosmic variance, see, e.g. \citealt{planck13-16}); $B$-mode polarization is the only window 
for measuring this signal in the CMB.
The amplitude of the gravitational wave background is
proportional to the expansion rate during inflation, and hence to the energy scale of the 
inflationary potential. Thus, a measurement of primordial $B$-mode polarization in the CMB
could potentially probe physics at energies approaching the Planck scale.

The CMB $B$ modes induced by gravitational lensing of primordial $E$ modes are both an
interesting signal in their own right and a potential contaminant to the inflationary gravitational-wave (IGW)
$B$ modes. In general, the signature of lensing in the temperature and polarization of the CMB 
can be used to reconstruct the projected gravitational potential $\phi$ between the observer and the 
last scattering surface \citep[e.g.,][]{seljak99}. 
The reconstructed potential is sensitive to the evolution of large-scale structure.
In particular, massive neutrinos affect the shape of this 
potential by streaming out of small-scale gravitational perturbations and damping
the growth of structure on these scales. A high-fidelity measurement of CMB lensing  
can in principle measure the sum of the neutrino masses 
\citep[e.g.,][]{abazajian15b}. The estimate of $\phi$ from CMB lensing
can also be used, in concert with a high-signal-to-noise map of the $E$ modes, to predict the
$B$-mode lensing signal.  
The predicted $B$-mode lensing signal can then be
cross-correlated with a direct $B$-mode measurement, as in e.g., \citet{hanson2013} (hereafter H13), 
or can be used to clean the lensing $B$ modes from a direct $B$-mode measurement, thereby improving sensitivity to 
IGW $B$ modes \citep{knox02,kesden02}.

Significant experimental progress has been made recently in the field of CMB $B$ modes.
The first detection of $B$ modes (H13)
was made in cross-correlation, using CMB data from 
SPTpol \citep{austermann12}---a polarization-sensitive receiver currently installed on the
10-meter South Pole Telescope \citep[SPT,][]{carlstrom11}---and cosmic infrared background (CIB)
data from \textit{Herschel}-SPIRE \citep{griffin10}.
Similar cross-correlation measurements have since been made 
\citep{polarbear2014c,vanengelen14b,planck15-15} 
using CMB data from POLARBEAR \citep{arnold09}, ACTPol \citep{niemack10}, and the \planck\ satellite.
The POLARBEAR team also published a measurement of the 
$B$-mode angular power spectrum \citep{polarbear2014b}
in which $\sim 2 \sigma$ evidence for the lensing signal was seen.
Finally, \citet{bicep2a} reported a strong detection of $BB$ power, including a component in excess 
of the \LCDM\ prediction for lensing $B$ modes at $\ell\sim80$.
The excess has been confirmed in yet deeper 150 GHz data on the same area of sky 
from the Keck Array experiment \citep{bicep2keck15}, but 
a recent joint analysis of BICEP2, Keck, and \planck\ data \citep{bicep2keckplanck15} has 
demonstrated that 
at least half of 
the excess is due to polarized emission from galactic dust 
and that the residual power is consistent with zero IGW signal.
Regardless of the interpretation of the excess $B$-mode signal, this analysis reported a 7$\sigma$ detection of lensing $B$ modes at $\ell \sim 200$, the tightest direct measurement of lensing $B$ modes to date.

In this paper, we present a measurement of the $BB$ spectrum in the multipole range 
$300 \le \ell \le 2300$,
estimated from 
100 \degs\ of data taken with SPTpol in 2012 and 2013.
The data used 
in this work have significant overlap with the data used to make the first detection of $B$ modes
in H13, but there are several key differences in the two data sets and analyses. First and most 
importantly, the analysis in H13 detected $B$ modes in SPTpol data by cross-correlating 
SPTpol $B$-mode maps with a predicted $B$-mode
template constructed using the $E$ modes measured with SPTpol and 
an estimate of $\phi$ from the CIB. Cross-correlation analyses have
the attractive property that any systematic effect present in only one of the data sets will not bias the 
result.
From a
systematics perspective, the $BB$ spectrum presented here is a much more
challenging measurement than the H13 cross-correlation measurement, and the
results presented here demonstrate that SPTpol cleanly measures the $B$ mode
power spectrum at angular scales of tens of arcminutes.
Furthermore, the $B$ modes measured in H13 are necessarily 
restricted to the lensing signal induced by the part of $\phi$ traced by the CIB (the CIB is expected
to have good but not perfect redshift overlap with the CMB lensing kernel, see, e.g., \citealt{holder13}).
The $BB$ spectrum presented in this work is sensitive to all contributing signals, including the 
full lensing signal, the IGW signal (though no detection of IGW $B$ modes is expected in the $\ell$
range probed in this work at the current levels of sensitivity), and any foreground contamination.
This work also contains 95 and 150 GHz data from both 2012 and early 2013, while H13 focused primarily on 150 GHz data and used only 2012 data.

This paper is structured as follows: 
Section 2 describes the telescope and receiver. Section 3 describes the observations used
in this analysis. Section 4 details the data reduction process from the raw, time-ordered
detector data to the stage of single-observation maps. Section 5 describes how we calculate
the power spectrum, including identifying and correcting for biases in the measured $BB$
spectrum, and presents the measured spectrum. We interpret the results in the context of
cosmology and other $B$-mode and lensing results in Section 6, and we conclude in 
Section 7.

\section{Telescope and Receiver}
\label{instrument}

The SPT is a 10-meter telescope with a wide field of view ($\sim 1$ square degree at 150\,GHz),
designed for conducting large-area surveys of fluctuations in the temperature
and polarization of the CMB. The telescope is described in detail in \citet{carlstrom11}, 
and further details about the optical design can be found in \citet{padin08}.

After five years of observation with the original SPT-SZ receiver, which was sensitive to the 
CMB intensity but not its polarization, the polarization-sensitive SPTpol receiver was 
installed in 2012.
SPTpol is equipped with 1536 polarization-sensitive
transition edge sensor (TES) bolometers, with
1176 detectors at 150\,GHz and 360 detectors at 95\,GHz.
The detectors in the two bands were designed and fabricated 
independently and are described in detail in \citet[][150\,GHz]{henning12} and 
\citet[][95\,GHz]{sayre12}.  
The 150\,GHz array is composed of seven detector modules,
each containing 84 pixels, all fabricated at the National Institute for Standards and Technology Boulder Laboratory.
Each module consists of a detector array behind a monolithic feedhorn array.  Incoming 
power is coupled through the feedhorns to an
orthomode transducer (OMT),
which splits the light into two orthogonal polarization states. 
The 95\,GHz array consists of 180 individually packaged
dual-polarization absorber-coupled polarimeters (a total
of 360 detectors) fabricated at Argonne National Laboratories.  
Each pair of 95\,GHz detectors is coupled to the telescope through machined
contoured feedhorns. The detectors in both observing bands
are read out using a digital frequency-domain multiplexer system with 
cryogenic superconducting quantum interference device (SQUID) amplifiers.

The focal plane is cooled to $\sim 250$~mK using a commercial pulse-tube
cooler and a three-stage helium refrigerator.  
The TES detectors are then biased to their operating point 
of $\sim 500$~mK. 
The entire receiver is maintained at $\sim 4$~K, similar to the
SPT-SZ receiver described in \citet{carlstrom11}. 
The secondary
optics (including the secondary mirror cryostat) are identical to those used for
SPT-SZ, with the exception of different heat-blocking filters near the prime focus.
For additional details on the SPTpol instrument design, see \citet{austermann12}.

\section{Observations}
\label{sec:obs}

The first eight months of observing with the SPTpol receiver (April-October 2012 and April 2013) were dedicated primarily to 
observations of a 100~\sqdeg\ patch of sky centered at right ascension 23h30m and declination $-55$ degrees.
We refer to this field as the SPTpol ``100d'' field to distinguish it from the 500~\sqdeg\ survey field,
for which observations began in May 2013.
The 95 and 150\,GHz data from the first year (2012) of observations of the 100d field 
were used in H13.
The 150\,GHz data from 2012 were also used to compute the $E$-mode power spectrum ($EE$) and the temperature-$E$-mode 
correlation spectrum ($TE$) in \citet[][hereafter C14]{crites14}.
\citet{story14} used the 150\, GHz data from both years of observation of this field to reconstruct the CMB lensing potential and analyze its power spectrum.
The analysis presented here
is the first to use the 95\, and 150\, GHz data on this field from both years, and 
the effective white noise levels are approximately 17 and 9 $\mu$K-arcmin in polarization at 95 and 150 GHz.

Some minor modifications to the instrument were made between the 2012 and 2013 
observing seasons that could affect the analysis in this work. At 150\,GHz, one detector
module was replaced, and the filters used to define the band edges were replaced in 
both bands. Both of these modifications most directly affect the 
shape and overall width of the observing bands, which can in turn affect beam shape
and absolute calibration. For this reason, we estimate beams and absolute calibration
independently for both seasons.

All observations of the 100d field used a constant-elevation
scan strategy, in which the telescope is slewed in azimuth back and forth once then stepped a
small amount in elevation, with the process repeated until the full elevation range is covered.
The azimuthal
scanning speed used in all observations was 0.48 degrees per second, corresponding to 0.28 
degrees per second on the sky at the mean elevation of the field, and 
the constant-velocity portion of the scans were between 8.8 and 10.75 degrees in azimuth.
The elevation step between scans was between 13 and 20 arcminutes.

Only one half of the azimuthal extent of the field is observed at one time, in a 
``lead-trail'' strategy that allows for ground contamination to be efficiently subtracted
when maps of the two field halves are differenced \citep[e.g.,][]{pryke09}. Since
no ground signal is detected in this analysis (see Section \ref{sec:jacks}) or in C14, 
we do not use a lead-trail differencing analysis in this work; instead, we simply combine
each pair of half-field maps into a map of the full field.

We refer to one pass of the telescope, either from left to right or from right to left across 
half of the field, as a ``scan,''
and we refer to a set of scans that cover an entire half-field as an ``observation.''
Each observation lasts 30 minutes, and there are a total of roughly 12,000 
observations in the 2012 and 2013 observing seasons, for a total of roughly 6000
individual-observation maps of the full field.

\section{Data Reduction: Time-Ordered Data to Maps}
\label{sec:data}
In their raw form, the time-ordered data for each observation used in this work consist 
of one vector of uncalibrated ADC counts for each detector, representing the 
current through the TES as a function of time, and two vectors representing
the detector pointing as a function of time. The first step in processing these data
into angular power spectra is to convert the time-ordered data into pixelized maps 
of the Stokes parameters $I$ (or $T$ for ``temperature''), $Q$, and $U$ on the sky.
The process used in this work for making maps from time-ordered data closely 
follows that of C14; we describe the stages of this process and point out any differences
with C14 below.

\subsection{Time-ordered Data Filtering}
\label{sec:filter}
The data from each detector in each observation are filtered prior to making maps for a number of reasons:
to remove modes that 
are strongly affected by low-frequency noise,  
to prevent aliasing of high-frequency
noise to lower frequencies when the data are binned into map pixels, and 
to eliminate potential contamination from detector coupling to the pulse-tube cooler
used to cool the receiver.
Data from each detector in
each scan are fit to a combination of a first-order polynomial (mean and slope) and a set of 
low-order Fourier modes (sines and cosines). The best-fit polynomial and Fourier modes
are removed, resulting in an effective high-pass filter. When maps are made from data
filtered in this way, the effect of this time-domain filtering is a high-pass filter along the 
scan direction (equivalent to right ascension for observations from the South Pole) 
with a cutoff at angular scales of roughly 
three degrees (or an effective multipole number in the scan direction of $\ell_x \sim 100$).\footnote{Throughout
this work, we use the flat-sky approximation to equate multipole number $\ell$ with 
$2 \pi |\mathbf{u}|$, where $\mathbf{u}$ is the Fourier conjugate of Cartesian angle on a 
patch of sky small enough that curvature can be neglected.} To avoid filtering 
artifacts around bright point sources, all sources above 50 mJy in unpolarized flux at 150\,GHz
are masked in the fitting process. 
An anti-aliasing low-pass filter is also applied to 
the data from each detector on each scan, resulting in a scan-direction low-pass filter
in the maps with a cutoff of $\ell_x \sim 6600$. This filter cutoff is 
set by the size of the pixels used in mapmaking, which is one arcminute for this analysis.
A frequency-domain filter with very narrow notches at the pulse-tube cooler frequency
(and the second and third harmonics of this frequency) is applied to the data from each detector 
over the entire observation. A total of $< 0.01$~Hz of bandwidth 
($<0.2\%$ of the $\ell_x<6600$ band) is removed with this filter, 
and we do not include its effect in the simulations of the filter transfer function described 
in Section \ref{sec:sims}.

The C14 analysis used slightly different filtering choices. 
The scan-direction 
high-pass filtering was accomplished in C14 using polynomial subtraction only (no Fourier modes); 
for this work, we found that the combination of polynomials and Fourier modes resulted in a lower level 
of spurious $B$-mode power created in the filtering step (see Section \ref{sec:additive} for details).
The anti-aliasing low-pass filter in C14 had a cutoff at $\ell_x \sim 10000$,
because the maps used in that analysis had 0.5-arcminute pixels.

\subsection{Relative Calibration}
\label{sec:relcal}
Before the data from all detectors are combined into $T$, $Q$, and $U$ maps, 
a factor is applied to the data from each detector to equalize the response to 
astronomical signal across each of the two detector arrays (95 and 150\,GHz). 
The process of measuring and monitoring this
relative calibration for each array is identical to the process described in C14
and used in earlier SPT analyses (see
\citealt{schaffer11}); we describe the process briefly here.

The relative calibration process involves regular observations of the galactic HII region RCW38 
and of an internal chopped blackbody source. 
A 45-minute observation of the galactic HII region RCW38 is conducted approximately once per day
(shorter observations for pointing reconstruction are conducted more frequently), while one-minute
observations of the internal source are conducted at least once per hour.
An effective temperature for the internal source is assigned individually for each detector
(by comparing to the season average of response to RCW38). This value times the 
response of each detector to the internal source in the observation nearest a CMB field observation
is used for relative calibration.  
We assume that RCW38 is unpolarized, and internal measurements suggest that it is $<1\%$ polarized.
In Section~\ref{sec:tqu} we discuss the potential for spurious $B$-mode polarization caused by small polarization in RCW38.

Drifts in the internal source temperature are accounted for by comparing the average 
source response across a detector module\footnote{Although the 95\,GHz detectors 
are fabricated individually, they are effectively grouped into four modules of 45 
dual-polarization pixels (90 detectors) each, 
based on the wiring of detectors to SQUIDs to SQUID readout boards.}
in each observation to the module average over the entire season. Drifts in atmospheric opacity are 
addressed similarly using average RCW38 responses for each module.

\subsection{Polarization Calibration}
\label{sec:polcal}
Before combining the data from all detectors
into $T$, $Q$, and $U$ maps, we must know the polarization properties of the detectors.
Each detector is
designed to be sensitive to linearly polarized radiation at a particular angle and insensitive
to radiation in the orthogonal polarization. The two numbers required to characterize each 
detector's polarization performance are the polarization angle $\theta$ 
(the angle of linearly polarized radiation at which the response of the detector is maximized)
and the polarization efficiency $\eta_p$ (a measure of the ratio of detector response to linearly 
polarized radiation at $\theta$ to radiation polarized at $\theta + \pi/2$).

These properties are measured for each detector in dedicated observations of a polarized calibration 
source during the Austral summer season. 
The calibration observations and the method by which 
$\theta$ and 
$\eta_p$ are derived from the data are described in detail in C14. We briefly summarize the 
observations and data reduction here.

The polarization calibrator consists of a chopped thermal source located behind two wire grid polarizers
(one at a fixed angle and one that can rotate), physically located 3 kilometers from the telescope.  
One dual-polarization pixel is pointed at the calibrator, and the rotating polarizer is stepped through
nearly 180 degrees while the response of the two detectors is monitored.
We fit the response as a function of rotating grid angle to a model including $\theta$ and $\eta_p$
as free parameters.
This procedure is repeated for all pixels, with multiple measurements per detector where possible. 
For detectors for which the measurements or model fits do not pass data quality cuts 
($\sim 25\%$ of the detectors used in this work),
we assign the median value measured from 
a subset of detectors, namely those that are on the same detector module, have the same nominal angle, 
and did pass data quality cuts.
We expect this to be a reliable substitution since these subsets of detectors were designed to have the same angles, and the successfully measured angles are consistent with the median angle.

The median statistical uncertainty on the detector polarization angles is 0.5$^\circ$ per detector, 
and the systematic uncertainty on these angles is estimated to be 1$^\circ$. 
The mean measured polarization efficiency is $98\%$, and the median statistical error on 
the efficiency is 0.7\% per detector. A correlated error in the estimation of all polarization angles will 
result in mixing between $E$ and $B$ modes, 
and this effect is addressed by the cleaning procedure described in Section \ref{sec:convclean}.

\subsection{Data Cuts}
\label{sec:cuts}
We flag and ignore time-ordered data from individual detectors on a per-scan basis and a 
per-observation basis using cut criteria that are nearly identical to those used in C14.
Briefly, we 
flag data from individual detectors on a per-scan basis  
based on detector noise and the presence of discontinuities in the data, including
spikes (generally attributed to cosmic rays) and sharp changes in DC level (generally attributed to 
changes in the SQUID operating point). Data from a particular detector is not used 
for the entire scan if either of these types of discontinuities is detected, or if the rms
of the data from that detector in that scan is 
greater than 3.5 times the median or less than 0.25 times the median. The median rms
is calculated across all detectors on a module. These cuts remove roughly five percent of the data.

Data from individual detectors are flagged at the full-observation level based on
response to the two types of calibration observations described in Section \ref{sec:relcal}
and noise in the frequency band of interest.
For a given CMB field observation, 
data from detectors with low signal-to-noise response to the closest observation of either RCW38 
or the internal blackbody source are excluded from the map for that CMB field observation.
The noise is calculated for each detector in each observation 
by taking the difference between left-going and right-going scans, 
Fourier transforming, and calculating the square root of the mean power between 0.3~Hz and 2~Hz
(corresponding to $400 \lesssim \ell \lesssim 2000$ in the scan direction, roughly the 
signal band of interest to this work).
Data from detectors with abnormally high or low noise in this band
are cut from an entire observation. We also enforce that both detectors in a dual-polarization
pixel pass all of these cuts; if one detector is flagged, then the pixel partner is flagged as well. 
We find empirically that this cut improves the low-frequency noise of the resulting maps, 
presumably by increasing the fidelity of the subtraction of unpolarized atmospheric signal.

\subsection{Map-making}
\label{sec:mapmaking}
After the cuts described in the previous section are applied, 
the filtered, relatively calibrated, time-ordered data from each detector are
combined into $T$, $Q$, and $U$ maps
using the pointing information, polarization angle, and a weight for each detector.
The procedure used in this analysis for polarized map-making is similar to earlier work, e.g., \citet{couchot99} and \citet{jones07}.
Briefly, for a single observation of the CMB field, 
weights are calculated for each detector based on detector polarization efficiency and 
noise rms---i.e., $w \propto (\eta_p / n)^2$, where $\eta_p$ is 
the polarization efficiency (one value per detector for all observations, based on the
polarization calibration observations described in Section \ref{sec:polcal}), and $n$ is the noise rms,
calculated as described in the previous section.
The contributions to the weighted estimates of the $T$, $Q$, and $U$ values for pixel $\alpha$
from detector $i$ are then

\begin{eqnarray}
\hat{T}^W_{i \alpha} &=& \sum_t A_{t i \alpha} \ w_i  \ d_{ti}  \\
\nonumber \hat{Q}^W_{i \alpha} &=& \sum_t A_{t i \alpha} \ w_i \ d_{ti} \ \cos{2 \theta_i} \\
\nonumber \hat{U}^W_{i \alpha} &=& \sum_t A_{t i \alpha} \ w_i \ d_{ti} \ \sin{2 \theta_i}
\end{eqnarray}

where $t$ runs over time samples, 
$d_{ti}$ is the value recorded by detector $i$ in sample $t$, $w_i$ is the weight 
for detector $i$ (defined above), $A_{ti \alpha}$ is the pointing operator encoding when detector $i$ was
pointed at pixel $\alpha$, and $\theta_i$ is the detector polarization angle, and the
hat implies measured quantities.

A 3-by-3 matrix representing the $T$, $Q$, and $U$ weights and the correlations between the three
measurements is created for each map pixel using this same information. 
The contribution to the weight matrix for the estimate of $T$, $Q$, and $U$ in pixel $\alpha$
from detector $i$ is

\begin{widetext}
\begin{equation}
W_{i \alpha} = \left( \begin{array}{ccc}
 w_i N_{i \alpha}  &   w_i N_{i \alpha} \cos{2 \theta_i} &  w_i N_{i \alpha} \sin{2 \theta_i}  \\
 w_i N_{i \alpha} \cos{2 \theta_i} &   w_i N_{i \alpha} \cos^2{2 \theta_i} &  w_i N_{i \alpha} \cos{2 \theta_i} \sin{2 \theta_i}  \\
 w_i N_{i \alpha} \sin{2 \theta_i} &   w_i N_{i \alpha} \cos{2 \theta_i} \sin{2 \theta_i} &  w_i N_{i \alpha} \sin^2{2 \theta_i}  \\
\end{array} \right)
\end{equation}
\end{widetext}

where $N_{i \alpha}$ is the number of samples in which detector $i$ was pointed
at pixel $\alpha$. The contributions to the weighted  $T$, $Q$, and $U$ estimates
and the weight matrix from each detector in a given observing band are summed, 
the weight matrix is inverted for each pixel, and the resulting 
3-by-3 matrix is applied to the weighted $T$, $Q$, and $U$ estimates to produce the
unweighted map:

\begin{equation}
\{\hat{T},\hat{Q},\hat{U}\}_\alpha = W^{-1}_\alpha \{\hat{T}^W_{\alpha},\hat{Q}^W_{\alpha},\hat{U}^W_{\alpha}\}.
\end{equation}

We make maps using the oblique Lambert azimuthal equal-area projection with 1-arcminute pixels.
This sky projection introduces small angle distortions, which we account for by rotating the 
$Q$ and $U$ Stokes components across the map to maintain a consistent angular coordinate system in this projection.
The maps are in units of $\mu$K$_{\mathrm{CMB}}$, the equivalent fluctuations of a 2.73\,K blackbody that would produce the measured deviations in intensity.

\subsection{Combining Single-observation Maps}
\label{sec:bundles}

The cross-spectrum analysis described in Section \ref{sec:xspec} assumes uniform
noise properties across all maps used in the analysis and all pixels in the maps. 
Maps made from a one half-hour observation (each) of the lead and trail
halves of the SPTpol 100d field have sufficiently non-uniform coverage that the 
cross-spectrum analysis performed on these single maps would be significantly
sub-optimal. We choose to combine the single-observation maps 
into 41 subsets or ``bundles'' and do the cross-spectrum analysis on these
bundles. As shown in \citet{polenta05}, the excess variance on a cross-spectrum 
of data equally split $N$ ways as compared to an auto-spectrum of the full data scales
as $1/N^3$, so the excess variance with 41 bundles is negligible.

In a process similar to the detector-level data cuts described in Section
\ref{sec:cuts}, we calculate the distribution of a number of variables over the
entire set of single-observation maps, and we only include in bundles the maps 
that lie within a certain distance of the median for all variables. As in C14, we 
cut maps based on the rms noise, the total map weight, the median map pixel weight,
and the product of median weight times the square of the rms noise. In this analysis, 
we additionally cut maps based on the number of contributing bolometers and 
on the rms deviation of the time-ordered data averaged across all detectors from a model of that 
average time-ordered data (assuming the primary signal is from the atmosphere).
We include only those maps that pass cuts at both 95 and 150 GHz.
A total of 4897 of the roughly 6000 individual observations of the full field are 
included, with approximately 120 observations included in each bundle.

\subsection{Absolute Calibration}
\label{sec:abscal}

The relative calibration process described in Section \ref{sec:relcal} results in single-observation
or bundled $T$ maps that are related to the true temperature on the sky by the filter
and beam transfer function and a single overall number, the absolute temperature calibration.
(The $Q$ and $U$ maps have an additional factor of the overall polarization efficiency, as discussed in Section~\ref{sec:abspolcal}.) The processes for estimating the beam and filter transfer function are described in Section \ref{sec:sims}; 
we obtain the absolute temperature calibration using
the same method used in C14. We briefly describe the method here; for details, see C14.

The absolute calibration for the 150\,GHz SPTpol data is derived by comparing full-season coadded temperature maps 
to temperature maps of the same field made previously with the SPT-SZ receiver, which were in turn calibrated through a 
comparison of the \citet{story13} SPT-SZ temperature power spectrum with the 
\citet{planck13-16} temperature power spectrum.
(The uncertainty on the SPT-SZ-to-\planck\ calibration is estimated to be $1.2\%$ in temperature.)
The 95\,GHz SPTpol data are then calibrated by comparing to the 150\,GHz SPTpol data.

We compare the 150\,GHz SPTpol maps used in this work to the 150\,GHz SPT-SZ maps of the same field
by creating cross-power spectra, using the same method we use for cross-spectra of SPTpol 
bundle maps, described in Section \ref{sec:xspec}.
Specifically, we calculate the ratio of the cross-spectrum between a full-depth SPTpol map and
a half-depth SPT-SZ map and the cross-spectrum of two half-depth SPTpol maps, corrected 
for the difference in beams between the two experiments. (The SPT-SZ maps are made
with the same scan strategy and filtering as the SPTpol maps, so the filter transfer function 
divides out of the ratio.) We repeat
this for many semi-independent sets of half-depth SPT-SZ maps
and use the distribution of the
cross-spectrum ratio to calculate the best-fit 150\,GHz absolute calibration and the uncertainty on this
calibration. 
A similar process is used to compare 95\,GHz SPTpol maps to 150\,GHz SPTpol maps and
to obtain the 95\,GHz absolute calibration.
In the power spectrum pipeline, we multiply all SPTpol maps by these calibration factors before 
calculating cross-spectra.

We calculate the absolute temperature calibration 
separately for 2012 and 2013 data because the shape and overall width of the
observing band changed slightly between the two seasons (see Section \ref{sec:obs} for
details). In all calculations, we treat the 150\,GHz SPTpol full- and half-depth maps as noise-free
compared to 95\,GHz SPTpol and 150\,GHz SPT-SZ full- and half-depth maps, because the 
roughly 3x higher noise in the latter two data sets dominates the cross-spectrum uncertainty budget.
The fractional statistical uncertainties on the calibration of SPTpol 150\,GHz data to SPT-SZ 150\,GHz data 
are $0.004$ for both seasons (2012 and 2013). In quadrature with the SPT-SZ-to-\planck\ uncertainty, 
this results in a total calibration uncertainty of $1.3\%$ at 150\,GHz for both seasons, highly correlated
between the two seasons. The statistical uncertainty on the comparison of SPTpol 95\,GHz data to
SPTpol 150\,GHz data is $< 0.002$ for both seasons, which contributes negligibly to the total uncertainty budget; thus, 
the total calibration uncertainty at 95\,GHz is also $1.3\%$ and nearly $100\%$ correlated with the
150\,GHz uncertainties.
All of these uncertainties are much smaller than the 
fractional uncertainty
with which we expect to measure the $BB$ spectrum, and they
will make a negligible contribution to the uncertainty on the power spectrum or any derived parameters.

\subsection{Absolute Polarization Calibration}
\label{sec:abspolcal}
Although the polarization angles and efficiencies are measured with a ground-based polarization source as described in Section~\ref{sec:polcal}, we allow for additional freedom in the normalization of the $Q/U$ maps.  This is to account for any mechanism that could alter the effective polarization calibration.  One such mechanism, the impact of electrical crosstalk on our effective polarized beam, is discussed in Section~\ref{sec:xtalk_beams}.
The polarization calibration is calculated by comparing the $EE$ spectrum measured in this work to the $EE$ spectrum from the best-fit \LCDM model for the \textsc{Planck+lensing+WP+highL} constraint in Table 5 of \cite{planck13-16}.  The calculation of the measured $EE$ spectrum mirrors that of the $BB$ spectrum as described in Section~\ref{sec:powspec}.  The polarization calibration factor for spectral band $m \in \{ 95~\rm{GHz}, 150~\rm{GHz} \}$ is calculated as 
\be
(P_{\rm{cal}}^{m})^2 = \sum_b{ \left( C_b^{EE, m \times m, \rm{theory}} / C_b^{EE, m \times m, \rm{data}} \right) w_b}
\ee
where $b$ is the $\ell$-bin index and $w_b$ are simulation-based, inverse-variance weights that sum to one.  We perform this calculation after applying the absolute temperature calibration described in Section~\ref{sec:abscal}.  We find $P_{\rm{cal}}^{95} = 1.015 \pm 0.024$ and $P_{\rm{cal}}^{150} = 1.049 \pm 0.014$, and we multiply the data polarization maps by these factors.  In Section~\ref{sec:systematics} we discuss the impact of the uncertainty of these calibration factors on the $BB$ power spectrum measurement.

\subsection{Apodization Mask}
\label{sec:apod}
We construct an apodization mask, $\mathcal{W}$, to downweight the high-noise regions at the boundary of the SPTpol coverage area and to mitigate  mode-coupling.  The apodization mask is constructed as follows.  Each map bundle has an associated weight array which is approximately inversely proportional to the square of the white noise level at each pixel of the temperature map.  We smooth each weight array with a $\sigma=10^\prime$ Gaussian kernel, threshold the smoothed array on 0.05 of its median value, and take the intersection of all thresholded arrays across all bundles.  We then pad the boundary of the intersection map with $5^\prime$ of zeros and smooth with a $\lambda=60^\prime$ cosine kernel.  We perform this process at 95 and 150 GHz, and the final apodization mask $\mathcal{W}$ is the product of the individual 95 and 150 GHz masks.

\subsection{Map-space Cleaning}
\label{sec:mapclean}

We perform two map-space cleaning operations on the bundle $\{T,Q,U\}$ maps.  These operations are performed on both the data maps and the simulation maps described in Section~\ref{sec:sims}.

First, to reduce our sensitivity to bright, emissive sources, we interpolate over regions of the map near all sources with unpolarized fluxes $S_{150}>50$ mJy.  For each source, we replace the values of the map within $r<6^\prime$ of the source with the median value of the map in an annulus defined by $6^\prime < r < 10^\prime$.

Second, we filter the maps to reduce our sensitivity to scan-synchronous signals.  
We see evidence for a polarized scan-synchronous signal with $\sim$5 $\mu \rm{K}$ rms at 95 GHz, and one with $\sim$0.6 $\mu \rm{K}$ rms at 150 GHz.
We clean these signals by removing from each map a template that is only a function of RA.  Due to SPT's polar location and constant-elevation scan strategy, RA is essentially equivalent to the scan coordinate, and any signal that depends only on the scan coordinate will depend only on RA.  We construct the scan-synchronous template for each map by averaging the map in bins of RA and  then smoothing this binned function with a $1^\circ$-in-RA Hann function.  The resulting one-dimensional template is expanded to a two-dimensional template and then subtracted from the original, two-dimensional map.

\section{Power Spectrum Analysis}
\label{sec:powspec}

In this section we describe the process by which the maps described in the previous section are reduced to $BB$ power spectra.

\subsection{Simulations}
\label{sec:sims}

A number of steps in the power spectrum analysis rely on mock maps, and here we describe the process by which the mock maps are generated.  

First, we generate noise-free sky maps that form the input to time-ordered-data simulations.  The input sky maps are Gaussian realizations of $\{T,Q,U\}$ generated in the \textsc{HEALPix} pixelized sphere format \citep{gorski05} with 0.43$^\prime$ ($n_{\mathrm{side}}=8192$) resolution.  The sky maps are generated using $\{TT,TE,EE,BB\}$ CMB power spectra computed using the \textsc{CAMB} Boltzmann code \citep{lewis00}.  The \LCDM cosmological parameters input to \textsc{CAMB} are taken from the \textsc{Planck+lensing+WP+highL} best-fit model in Table 5 of \cite{planck13-16}.  There is no tensor power ($r=0$), and no foreground power.  There are two sets of input sky maps: 

\begin{itemize}
\item{$\mathbf{TEB}$ - These sky maps use all of the non-zero \LCDM CMB power spectra: $\{TT,TE,EE,BB\}$.  The $BB$ spectrum is due to gravitational lensing of $E$-mode polarization.}
\item{$\mathbf{TE}$ - These sky maps are identical to those described above, including identical random seeds, but do not include $B$-mode polarization power: $C^{BB}_\ell=0$.}
\end{itemize}
We generate 100 realizations of each set of input sky maps and convolve them with an azimuthally-symmetric beam function.  The beam function is measured using observations of planets and PKS 2355-534, the brightest source in the 100d field at these observing frequencies.  We use four distinct beam functions, one for each combination of spectral band and year of observations, to generate input sky maps for each spectral band and year of observation.

Next, we use the SPTpol pointing information to generate time-ordered data representing mock observations of the input sky maps.    Before reducing the time-ordered data to maps, we simulate the effect of ``crosstalk'', the mixing of time-ordered data between different detectors.  We have used observations of the galactic HII region RCW38 to measure percent-level crosstalk between some detectors.  The crosstalk is believed to originate in the readout electronics.  We define a crosstalk matrix $V_{ab}$ to denote the coupling between detectors $a$ and $b$, and we use $V_{ab}$ to mix the simulated time-ordered data of different detectors.  The main effect of crosstalk for the $BB$ analysis presented here is  \TtoB~leakage, which we discuss further in Section~\ref{sec:additive}.  After introducing the effect of crosstalk, the simulated time-ordered data are reduced to bundle maps in a manner that matches the reduction of the real data as described in Section~\ref{sec:data}.

Finally, we add realistic noise to the simulated bundle maps using differences of the data maps.  
We generate a single ``realization'' of noise for a specific simulated map bundle by combining a random half of its $\sim$120 constituent data maps into one coadded map, the remaining half into another coadded map, and differencing the two.  The resulting difference has no true sky signal and should be statistically representative of the noise in this particular map bundle, modulo one small difference.  When making the normal map bundles we combine the two halves with their respective weights, but when making the noise bundles we combine with equal weights.  We estimate that this results in the power in the noise bundles being $\sim$0.1\% larger than the true noise power.

We generate 200 realizations of noise for each bundle in this manner.  The constituent data maps are differenced differently in each noise realization.  In total there are 200 realizations of noisy map bundles.  Each of the 100 input sky realizations is used twice, while each of the 200 noise realizations is used once.

\subsection{Constructing $E$ and $B$}

Here we describe the process by which $E_{\pmb{\ell}}$ and $B_{\pmb{\ell}}$, the harmonic-space representations of the $E$-mode and $B$-mode polarizations, are constructed from the real-space $\{Q,U\}$ maps.
First we address small angle distortions introduced by our use of the Lambert azimuthal equal-area projection.  We account for these distortions by rotating the $Q$ and $U$ Stokes components across the map to maintain a consistent angular coordinate system.

To construct the harmonic-space representation of the $E$-mode polarization we use the standard convention \citep{zaldarriaga01}
                                                                                                                     
\be
E_{\pmb{\ell}} = Q_{\pmb{\ell}} \cos{2\phi_{\pmb{\ell}}} + U_{\pmb{\ell}} \sin{2\phi_{\pmb{\ell}}}
\ee
where $\phi_{\pmb{\ell}}$ is the azimuthal angle of $\pmb{\ell}$, $P_{\pmb{\ell}} \equiv \mathcal{F}\{P \mathcal{W}\}$ is the Fourier transform of each apodized map $P \in \{Q,U\}$, and $\mathcal{W}$ is the apodization window.

To construct the harmonic-space representation of the B-mode polarization we use the $\chi_B$ method described in \citet{smith07b}.  In this method, an intermediate $\chi_B$ map is constructed as the sum of convolutions of the unapodized $Q$ and $U$ maps.  The kernels for these convolutions are compact in real space; each pixel in the $\chi_B$ map is a curl-like linear combination of all $Q/U$ pixels within the surrounding square of $5\times5$ pixels.  Finally, we construct $B_{\pmb{\ell}}$, the harmonic-space representation of the B-mode polarization, 

\be
B_{\pmb{\ell}} = \left( (\ell-1)\ell(\ell+1)(\ell+2) \right)^{-\frac{1}{2}} \mathcal{F}\{ \chi_B \mathcal{W} \}
\ee
where $\mathcal{F}\{ \chi_B  \mathcal{W} \}$ is the Fourier transform of the apodized $\chi_B$ map, and the $\sim\ell^{-2}$ factor accounts for the second-order spatial derivative present in the curl-like linear combination of the $Q$ and $U$ maps.

\label{sec:eb}

\subsection{Cross-spectra}
\label{sec:xspec}
We employ a cross-spectrum pseudo-$C_\ell$ approach to estimate the $BB$ power spectrum, as described below.
The first step is to define the two-dimensional cross-spectrum between two CMB fields $(\alpha, \beta) \in \{T,E,B\}$, each coming from a spectral band $(\nu_m, \nu_n) \in \{95~\rm{GHz}, 150~\rm{GHz} \}$:

\be
\widehat{C}^{\alpha^{m} \beta^{n}}_{\pmb{\ell}} \equiv 
\frac{1}{n_{\rm{pairs}}}
\sum_{(i,j), i \ne j} 
\mathrm{Re}{ \left\{ (\alpha^{m,i}_{\pmb{\ell}})^{*} \beta^{n,j}_{\pmb{\ell}} \right\} }
\ee
where $(i,j)$ are indices of distinct map bundles.  These quantities are defined on a two-dimensional Fourier grid with resolution $\delta \ell=25$.  The next step is to calculate one-dimensional bandpowers.  These are the weighted sums of the two-dimensional spectra within a set of $\ell$-bins $\{b\}$,

\be
\widehat{C}^{\alpha^{m} \beta^{n}  }_{b} \equiv 
\sum_{\pmb{\ell} \in b} \widehat{C}^{\alpha^{m} \beta^{n}  }_{\pmb{\ell}} H^{mn}_{\pmb{\ell}}.
\ee
The bandpowers are defined on five $\ell$-bins, each $\delta \ell=400$ wide, running from $\ell_{\mathrm{min}}=300$ to $\ell_{\mathrm{max}}=2300$.  The two-dimensional Fourier weight $H^{mn}_{\pmb{\ell}}$ is defined as 

\be
H^{mn}_{\pmb{\ell}} \equiv W^{mn}_{\pmb{\ell}} Z^{mn}_{\pmb{\ell}}
\ee
where $W^{mn}_{\pmb{\ell}}$ is a Wiener filter optimized for detecting a \LCDM~lensed $BB$ power spectrum and $Z^{mn}_{\pmb{\ell}}$ is a binary mask.  The Wiener filter is defined as 

\be
W^{mn}_{\pmb{\ell}} \equiv 
\frac{\langle  
\widehat{C}^{B^{m} B^{n}, \mathbf{TEB}}_{\pmb{\ell}} -  \widehat{C}^{B^{m} B^{n}, \mathbf{TE}}_{\pmb{\ell}}
 \rangle_{\mathrm{sims}}}
{\mathrm{Var} \left \{ \widehat{C}^{B^{m} B^{n}, \mathbf{TE}}_{\pmb{\ell}} \right \}_{\mathrm{sims}} }
\ee
where both the mean and variance are taken across the set of 200 noisy simulations.  The numerator is the mean two-dimensional power spectrum of the signal of interest, the lensed $BB$ power spectrum, and the denominator is the two-dimensional variance in the absence of this signal.  We note that this filter is optimized to reject the hypothesis of no lensed $BB$ power, while the filter optimized to constrain the amplitude of the lensed $BB$ power would use $\mathbf{TEB}$ simulations in the denominator.  Given that the data presented here are noise-variance-limited, the two filters are quite similar.  Both the numerator and denominator are smoothed by a Gaussian kernel with full-width at half-maximum FWHM$_\ell=150$ prior to taking their ratio.

The binary mask $Z^{mn}_{\pmb{\ell}}$ is used to mask out modes satisfying $| \ell_x |<175$ or $| \ell_y |<150$ for any spectrum involving 95 GHz data.  This mask is necessary to pass some of the jackknife tests described in Section~\ref{sec:jacks}.

\begin{figure*}
\begin{center}
\includegraphics[width=0.70\textwidth]{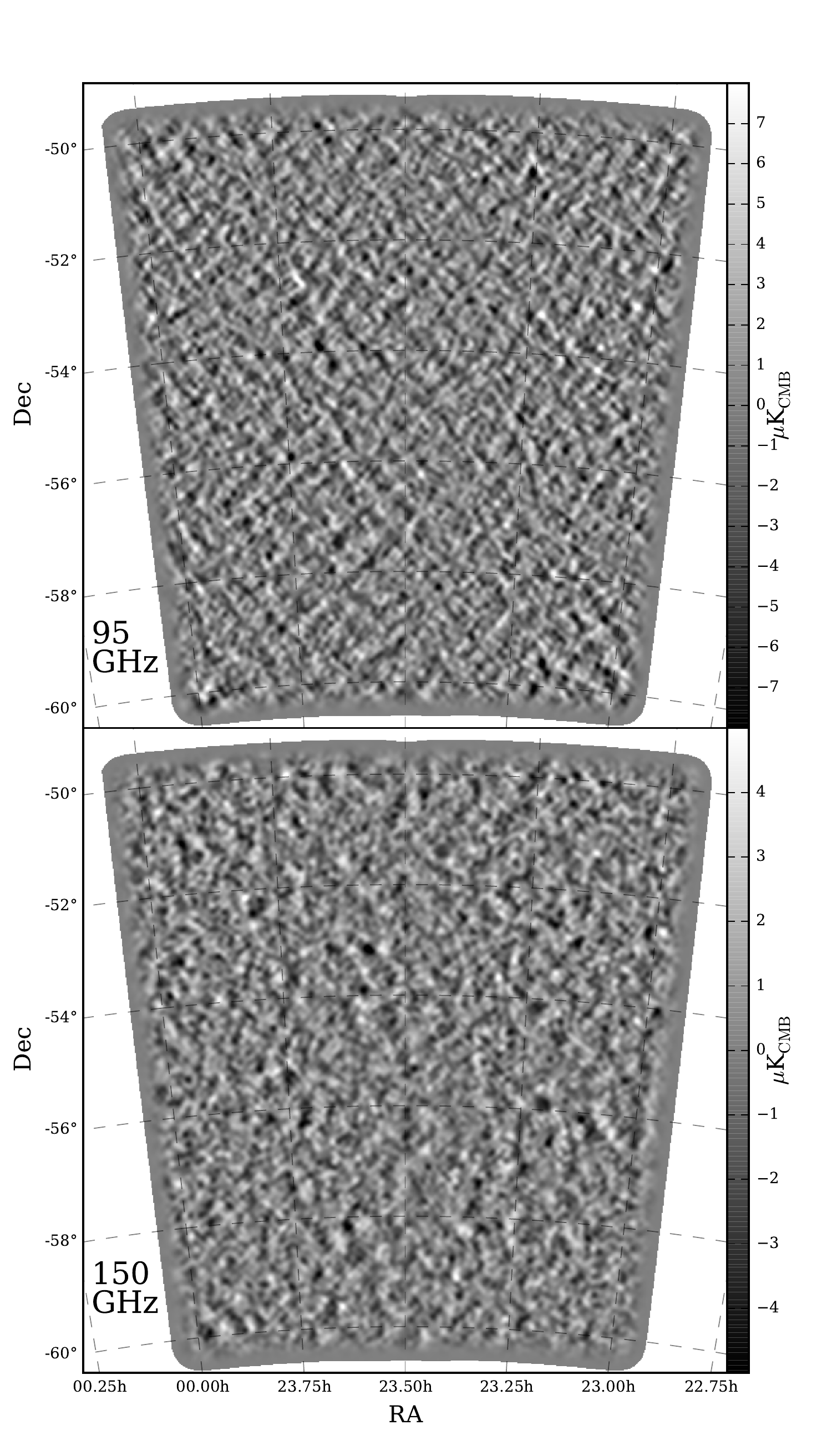}
\end{center}
\caption{The real-space representation of the $B$-modes analyzed in this work, at 95 GHz (top) and 150 GHz (bottom).  The $B$-modes have been filtered by the two-dimensional Fourier weights $\sqrt{H_{\pmb{\ell}}}$, as defined in Section~\ref{sec:xspec}, and multiplied by the apodization mask.
The crosshatch pattern faintly visible in the 95 GHz map is due to 
$H_{\pmb{\ell}}$ being zero near $\ell_x=0$ and $\ell_y$=0.
These maps are noise-dominated on all angular scales.}
\label{fig:bmap}
\end{figure*}

We note that while $H^{mn}_{\pmb{\ell}}$ has been defined to optimize sensitivity to the lensed $BB$ spectrum, we simply use the same weight for the other CMB field combinations ($EB$, $EE$, etc.), as the $BB$ spectrum is the focus of interest for this work.  In Figure~\ref{fig:bmap} we show the real-space representation of the $B$-modes analyzed in this work, namely $B^{m} =  \mathrm{Re} \{ \mathcal{F}^{-1}\{ \sqrt{H^{mm}_{\pmb{\ell}}} B^{m}_{\pmb{\ell}} \} \}$.  Each map $B^{m}$ has been multiplied by the apodization mask $\mathcal{W}$ for the purpose of visualization.

\subsection{Convolutional Cleaning}
\label{sec:convclean}
Given that the CMB temperature and $E$-mode polarization signals are expected to contain significantly more power than the $B$-mode polarization signal, the potential for leakage from $T$ or $E$ into $B$ is a central issue for any $BB$ analysis.  The leakage from a number of instrumental effects --- gain mismatch, differential pointing, differential beam ellipticity, incorrect polarization angles, crosstalk, etc. --- appears (to first order) as a spatially-local leakage of the $T$ or $E$ skies into $B$.
In other words, the leakage appears as a convolution of $T$ or $E$ into $B$.  
Here we describe a process, ``convolutional cleaning'', by which we reduce the impact of convolutional $(T,E) \rightarrow B$ leakage.  The basic idea is to use the CMB $T$ and $E$ modes themselves to estimate and project out convolutional leakage for $O(100)$ convolution kernels per spectral band.  We note that, after applying this cleaning, we are largely insensitive to any potential parity-violating cosmological $TB$ or $EB$ signal.

We consider leakage of $T$ or $E$ into $B$ that 
is described by a real-space convolution kernel, or equivalently, a multiplication in two-dimensional Fourier space by a complex coupling function $f_{\pmb{\ell}}$.  For example, the leakage from $T$ to $B$ for a particular coupling function $f_{\pmb{\ell}}$ looks like

\be
B_{\pmb{\ell}} = A~f_{\pmb{\ell}} T_{\pmb{\ell}}.
\ee
The strength of the leakage in the data is set by the parameter $A$, and we estimate $A$ in each $\ell$-bin $b$ for each coupling function.  In other words, we make the approximation that the amplitude of a particular form of leakage is constant over one $\ell$-bin.  There is still freedom for $\ell$-dependence of the leakage because the amplitude is allowed to vary between $\ell$-bins.

We use coupling functions of the form

\be
f^{k \theta}_{\pmb{\ell}} = \cos(k\phi_{\pmb{\ell}} + \theta)~i^k
\ee
where $k$ is a small integer $k \in \{0,1,...,k_{\mathrm{max}} \}$, $\phi_{\pmb{\ell}}$ is the azimuthal angle in two-dimensional Fourier space, $\theta \in \{0,\pi/2 \}$, and the $i^k$ factor ensures that $f$ corresponds to a purely real (rather than imaginary) real-space kernel.  We use this form because the coupling functions for well known types of leakage can be expressed as linear combinations of couplings with the form described above.  For example, differential gain has a $TB$ coupling $f_{\pmb{\ell}} \propto \cos(2\phi_{\pmb{\ell}})$, differential pointing has a $TB$ coupling $f_{\pmb{\ell}} \propto i \ell \left(\cos(\phi_{\pmb{\ell}})+\cos(3\phi_{\pmb{\ell}}) \right)$, and a global polarization angle offset has an $EB$ coupling $f_{\pmb{\ell}} \propto 1$.  We emphasize that we do not assume a particular physical mechanism (such as differential pointing or differential ellipticity) for this leakage.  Rather, we assume that the leakage is varying slowly in $\ell$, and has an azimuthal dependence described by $\cos(k\phi_{\pmb{\ell}} + \theta)$ for small integer $k$.
We use $k_{\mathrm{max}}=4$, which allows for the cleaning of any of the types of leakage described above.
We find that the main result of this work, the constraint on the amplitude of lensing $B$-mode power, is insensitive to the particular choice of $k_{\mathrm{max}}$; the lensing amplitude shifts by $\sim$0.1$\sigma$ when trying $k_{\mathrm{max}} \in \{2,4,6\}$.

The linear estimate of the amplitude $A$ of a particular form of leakage $f^{k\theta}_{\pmb{\ell}}$ in $\ell$-bin $b$ of the data $B^{m}_{\pmb{\ell}}$ for spectral band $m$ is 

\be
\hat{A}^{k \theta X m}_b = \frac{\sum_{\ell\in b}{ \mathrm{Re} \{ (B^{m}_{\pmb{\ell}})^{*} X_{\pmb{\ell}} f^{k \theta}_{\pmb{\ell}} \}  H^{mm}_{\pmb{\ell}} } }
{ \sum_{\ell\in b}{ | X_{\pmb{\ell}} f^{k \theta}_{\pmb{\ell}} |^2 H^{mm}_{\pmb{\ell}} } }
\ee
where $X \in \{T,E\}$, and the weight $H^{mm}_{\pmb{\ell}}$ is defined in Section~\ref{sec:xspec}.  The numerator is linear in the data $B^m_{\pmb{\ell}}$, and the denominator is used to normalize the estimate.  
Because the 150 GHz data are deeper than the 95 GHz data, we use the 150 GHz $T_{\pmb{\ell}}$ and $E_{\pmb{\ell}} $ modes to estimate the leakage in the 95 and 150 GHz $B_{\pmb{\ell}}$ modes.  
We use a cross-spectrum approach to estimate both $  \mathrm{Re} \{ (B^{m}_{\pmb{\ell}})^{*} X_{\pmb{\ell}} f^{k \theta}_{\pmb{\ell}} \}$ and $| X_{\pmb{\ell}} f^{k \theta}_{\pmb{\ell}} |^2$.  For our baseline value of $k_\mathrm{max}=4$, there are a total of 18 leakage modes per $\ell$-bin, or 90 modes for all five $\ell$-bins, per spectral band.

Ideally we would use these estimates of the leakage amplitudes to remove from the data $B^m_{\pmb{\ell}}$ 
the inferred leaked $B$-mode,

\be
B^{\mathrm{leak},m}_{\pmb{\ell}} = \sum_{k \theta X b} \hat{A}^{k \theta X m}_{b} X_{\pmb{\ell}}  f^{k \theta}_{\pmb{\ell}} q_{b  \pmb{\ell}}
\ee
where the two-dimensional binning operator $q_{b \pmb{\ell}}$ is one for $\pmb{\ell}\in b$ and zero otherwise.
In practice, however, due to the non-orthogonality of the different modes being removed, the cleaning process does not converge after one iteration.  We address this issue using a damped, iterative scheme.  In each iteration we estimate the amplitude of each form of leakage; subtract a damped version of the inferred leaked mode, $0.2 B^{\mathrm{leak},m}_{\pmb{\ell}}$, from $B^m_{\pmb{\ell}}$; and repeat for all forms of leakage.  We iterate over these steps 20 times, at which point the process has converged.  We use the accumulated estimate for each leakage amplitude to construct the final inferred leakage mode, $B^{\mathrm{leak},m}_{\pmb{\ell}}$.  Finally, we subtract $B^{\mathrm{leak},m}_{\pmb{\ell}}$ from $B^{m,i}_{\pmb{\ell}}$ for each bundle $i$.  To be clear, the same $B^{\mathrm{leak},m}_{\pmb{\ell}}$ is subtracted from all bundles.  In other words, all two-dimensional $B_{\pmb{\ell}}$ that appear in this work have been cleaned in this way, with one exception: the Fourier weights $H_{\pmb{\ell}}$ are calculated using $B_{\pmb{\ell}}$ that have not been cleaned.

We find that the convolutional cleaning process strongly suppresses the one form of convolutional leakage that we expect to be present in the SPTpol data: crosstalk-induced leakage.  We have used our simulated maps to determine that convolutional cleaning reduces the additive bias to the $BB$ spectrum introduced by crosstalk by factors of $\sim$20-100.
The cost of this improvement, however, is the introduction of a small, additive noise bias, as discussed below in Section~\ref{sec:additive} and in the Appendix.

\subsection{Unbiased Spectra}
\label{sec:unbiased}
Here we describe the process by which the raw $BB$ bandpowers described in Section~\ref{sec:xspec} are processed into unbiased bandpowers.  The raw bandpowers are subject to additive and multiplicative biases, and we correct for these biases using simulated bandpowers.

\subsubsection{Additive Bias}
\label{sec:additive}
Our strategy for dealing with additive bias --- any $BB$ power that is present in the absence of a true $B$-mode polarization signal --- is straightforward.  We measure the mean $BB$ power present in noisy, simulated bandpowers generated using $\mathbf{TE}$ input skies (i.e., no true $BB$ power), and we subtract this bias from the data bandpowers and the simulated $\mathbf{TEB}$ bandpowers.  This subtraction is performed prior to correcting for any multiplicative bias.  The additive biases can to some degree be organized as follows.

\begin{figure}
\includegraphics[width=0.55\textwidth]{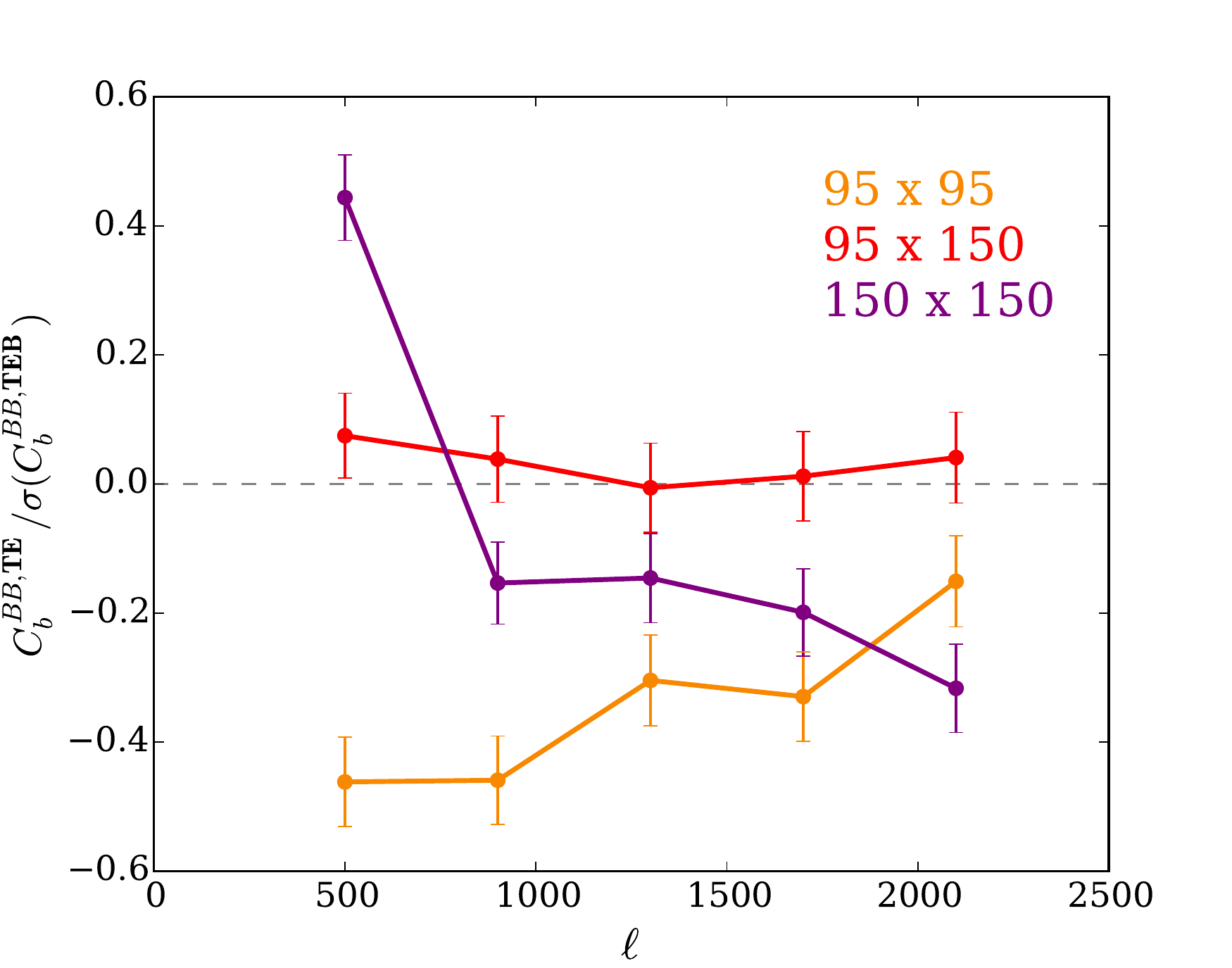}
\caption{The net additive bias to the three $BB$ spectra as a fraction of the diagonal statistical uncertainty.  Both of these quantities are determined using noisy simulated data.  The biases are $<0.5\sigma$ and are subtracted from the measured spectra.
}
\label{fig:te_bias}
\end{figure}

\begin{itemize}

\item{\emph{\EtoB~from geometry and filtering} - 
Imperfect separation of $E$ and $B$ on a small area of sky and the high-pass filtering of the time-ordered data result in \EtoB~leakage.  The resulting bias in $BB$ is approximately +$0.6 \sigma$ in the lowest $\ell$-bin of the $150 \times 150$ spectrum, and smaller than $0.1 \sigma$ in other $\ell$-bins of this spectrum or any $\ell$-bin of the other spectra. }

\item{\emph{$(T,E) \rightarrow B$ from crosstalk} - 
The electrical crosstalk between detectors results in $(T,E) \rightarrow B$ leakage, predominantly \TtoB.  Although the raw crosstalk-induced additive bias is as high as $1.0\sigma$ in the lowest $\ell$-bin of 150$\times$150, convolutional cleaning reduces this to $<0.1\sigma$.}

\item{\emph{Negative noise bias from convolutional cleaning} - 
We find that the convolutional cleaning step introduces a negative bias.  
By analyzing two sets of simulations which have convolutional cleaning turned on and off, we can isolate this effect.
We find that its magnitude and shape are in broad agreement with a simplified analytic calculation given in the Appendix.
The amplitude of the bias is approximately $-0.4\sigma$ in the lowest two $\ell$-bins of the $95\times95$ and $150\times150$ spectra, smaller elsewhere, and approximately zero at $95\times150$.
The convolutional cleaning process has strongly attenuated biases arising from convolutional leakage (crosstalk, beam systematics, etc.) in exchange for a small noise bias that we can characterize effectively perfectly.}

\end{itemize}
The additive biases are shown in Figure~\ref{fig:te_bias}.  We subtract these biases from the measured spectra, and we address systematic uncertainties associated with the additive biases in Section~\ref{sec:systematics}.

\subsubsection{Multiplicative Bias}
\label{sec:multiplicative}

After removing the additive bias we correct for the multiplicative bias.
 The bandpowers $\widehat{C}_b$ are a biased estimate of the true binned sky power, $C_{b^\prime}$, due to effects such as time-ordered data filtering, beam smoothing, finite sky coverage, and mode-mode mixing from the source and apodization mask.
The biased and unbiased estimates are related by

\begin{equation}
\widehat{C}^{B^{m} B^{n}}_{b}
\equiv K^{ B^{m} B^{n}  }_{ bb^\prime } C^{ B^{m} B^{n} }_{b^{\prime}}
\end{equation}
where the $K$ matrix accounts for the effects of the instrumental beam and time-ordered data filtering and the application of the apodization mask 
($\mathcal{W}$).
$K$ can be expanded as
\begin{equation}
\label{eqn:kdef}
K_{bb^\prime}=P_{b\ell}\left(M_{\ell\ell^\prime}[\mathcal{W}]\,F_{\ell^\prime}\right)Q_{\ell^\prime b^\prime}.
\end{equation}
where $F_{\ell}$ is the effective transfer function from the beam and the filtering of the time-ordered data, $P_{b\ell}$ is the binning operator and $Q_{\ell^\prime b^\prime}$ its reciprocal \citep{hivon02}.
The mode coupling kernel $M_{\ell\ell^\prime}[\mathcal{W}]$ accounts for the mixing of power between bins due to the apodization mask.
The mode coupling kernel is calculated analytically, as described in the appendix of C14.  The calculation corrects only for $B \rightarrow B$ coupling and effectively assumes that there is no \EtoB~coupling.  The latter is of course not true, but we correct for \EtoB~leakage by subtracting off an additive bias, as described in Section~\ref{sec:additive}.  As in C14, the effective transfer function $F_\ell$ is calculated by comparing the mean simulated spectra to the theory spectrum input to the simulations.  The calculation is iterative and converges after two iterations.

\subsection{Bandpower Covariance}
\label{sec:covariance}

We approximate the covariance between $BB$ bandpowers as completely diagonal, and we take the variance of bandpowers from the set of 200 noisy simulations as our estimate of the diagonal variances.  There are two sets of covariances: those that use $\mathbf{TE}$ input skies and thus naturally account for the noise variance and the variance from $T$ and $E$ leakage, and those that use $\mathbf{TEB}$ input skies, which additionally account for variance from true $BB$ power.  The $\mathbf{TE}$ variances are useful for rejecting the $C^{BB}=0$ hypothesis.  The $\mathbf{TEB}$ variances are only slightly larger than the $\mathbf{TE}$ variances, at most 20\% larger, implying that noise power dominates over signal power, as expected.

We have used the noisy simulated bandpowers to assess the validity of the diagonal covariance approximation.  We attempted to measure the covariance between off-diagonal spectral combinations e.g., (95 GHz $\times$ 95 GHz) and (95 GHz $\times$ 150 GHz), or between neighboring $\ell$-bins.  In each case, the distribution of covariance estimates is consistent with zero.  This is not surprising given that the covariance is dominated by noise power and the noise is expected to be uncorrelated between spectral bands and between $\ell$-bins.  We have also explicitly tested the importance of the off-diagonal covariance terms at (95$\times$95)$\times$(95$\times$150) and (150$\times$150)$\times$(95$\times$150).  We estimated these terms using an analytic approximation, altered the covariance matrix accordingly, and found that the main result of this work, the constraint on the amplitude of lensing $B$-mode power, changed by $\sim$0.1$\sigma$.  For simplicity, we do not include these off-diagonal terms.

We note that our simulation input skies contain Gaussian realizations of lensed $BB$ power and therefore do not account for the $\sim$20\% inter-$\ell$-bin correlation present for non-Gaussian, truly-lensed $B$-modes \citep{benoitlevy12}.  Given that the non-Gaussian structure leads to a 20\% bin-bin correlation on a source of power (lensed $BB$ power) that itself contributes only 10-20\% of the bandpower variance, we ignore the non-Gaussian contribution in our $\mathbf{TEB}$ variances.

Finally we note that, because our simulations were free of foreground power, we have slightly underestimated the $BB$ variance.  Power from randomly distributed, polarized point sources should be negligible; even the upper limit of the Poisson powers considered in Section~\ref{sec:interpretation} are $<1\%$ of the data noise power.  Variance from polarized galactic dust should also be small compared to the noise power.  
For our nominal dust model described in Section~\ref{sec:bbsig}, the dust power is approximately 1\% of the noise power in the lowest $\ell$-bin of the 150 GHz $\times$ 150 GHz spectrum, and smaller elsewhere.

\subsection{Bandpower Window Functions}
\label{sec:bpwf}
We compute bandpower window functions to allow the measured bandpowers to be compared to theoretical power spectra.
The window functions, 
$w^b_\ell / \ell$, 
are defined through the relation

\begin{equation}
C_b = (w^b_\ell / \ell) C_\ell.
\end{equation}

Following the formalism described previously in Section~\ref{sec:multiplicative}, this can be rewritten as

\begin{equation}
C_b = (K^{-1})_{bb'} P_{b' \ell'} M_{\ell' \ell} F_\ell C_\ell,
\end{equation}

which implies that

\begin{equation}
w^b_\ell / \ell = (K^{-1})_{bb'} P_{b' \ell'} M_{\ell' \ell} F_\ell.
\end{equation}

\subsection{Null Tests}
\label{sec:nulls}

In this section we test the internal consistency of the data by considering two types of null tests.  First we consider the $TB$ and $EB$ spectra.
Second we consider a suite of ``jackknife'' tests that are sensitive to potential instrumental systematics.

\subsubsection{$TB$ and $EB$ spectra}
\label{sec:tbeb}

Although we are interested primarily in the $BB$ power spectrum, here we consider the other spectra that contain the $B$-mode polarization field, namely the $TB$ and $EB$ spectra.  We do not expect a cosmological signal in these spectra.

Typically one would compare the $TB$ and $EB$ data spectra to the null hypothesis, namely $C^{TB}=C^{EB}=0$.  Here, due to the added complexity of the convolutional cleaning step, we do not necessarily expect the simulated $TB$ and $EB$ spectra to have zero mean, and indeed we find that several of the simulated bandpowers have negative mean values.  Nevertheless, we can test whether the data spectra are consistent with the distribution of simulated spectra.  We find that the $TB$ and $EB$ data spectra at $95 \times 95$, $95 \times 150$, and $150 \times 95$ are all consistent with the distribution of simulations.  
However, the $TB$ and $EB$ data spectra at $150 \times 150$ are systematically lower than the mean of the simulated spectra.  
$TB$ is low by [-2.3, -2.1, -1.8, -1.4, -1.4]$\sigma$ in the different $\ell$-bins, while $EB$ is low by [-1.5, -1.5, -3.0, -0.6, -1.5]$\sigma$.

While we cannot offer an explanation for this difference between the data and the simulations, we argue that it is not significant for the main focus of this work, the $BB$ spectrum.  More specifically, we use the difference between the data and simulation spectra to estimate the resulting spurious $BB$, and find that it is small compared to our statistical uncertainties.  For example, for $TB$ the difference between the data and simulations is $\delta C_b^{TB} \equiv C_b^{TB,\rm{data}} - \langle C_b^{TB,\rm{sims}}\rangle$, and the estimate of the associated spurious $BB$ power is $C_b^{BB} = (\delta C_b^{TB})^2/C_b^{TT}$.  We find the spurious $BB$ estimated from either $TB$ or $EB$ to be very small, less than 0.2\% of the expected \LCDM lensed $BB$ spectrum.

\subsubsection{Jackknives}
\label{sec:jacks}

We perform a suite of ``jackknife'' tests to further validate the consistency of the data.  In each jackknife test the bundle maps are sorted according to a metric designed to trace a potential systematic effect, and ``jackknife maps'' are constructed by differencing high-metric bundle maps with low-metric bundle maps.  The resulting jackknife maps should not contain sky signal and should have a power spectrum, the jackknife spectrum, consistent with noise.  We construct the jackknife spectrum by taking the mean cross-spectrum among the jackknife bundle maps.  We note that we do not apply the convolutional cleaning step to the bundle maps going into the jackknife maps.  The convolutional cleaning step would have no effect, as it would remove the same modes identically from each of the maps prior to taking their difference.

We perform four jackknife tests.
\begin{enumerate}

\item  Left-Right: Jackknife maps are made by differencing data from left-going and right-going scans.  This tests for any power or systematic effect that is present more strongly in one scan direction, such as spurious power caused by the step in telescope elevation at the end of each left-right scan pair.

\item Ground: Jackknife maps are made by dividing the maps into two subsets thought to have different susceptibility to ground contamination.  We use the same azimuthal range metric used in SPT-SZ power spectrum analyses such as \citealt{shirokoff11}.  This tests for spurious power from nearby buildings or other sources of contamination that are fixed to the ground.

\item 1st half-2nd half: Jackknife maps are made by differencing data from the first and second halves of
the set of chronologically ordered map bundles.
This tests for any systematic associated with the changes made to the SPTpol receiver between 2012 and 2013, systematics associated with small changes to the scan strategy made in late 2012, or any other slowly-varying systematic.

\item Visual Inspection: Jackknife maps are made by differencing map bundles that show visually-identified anomalies (e.g., faint stripes or spots) and those that do not.  This tests for systematic power associated with these features.

\end{enumerate}

We use a slightly different notation for the jackknife bandpowers than is used for the main bandpowers.  Each jackknife bandpower is $C^{fsj}_b$, where $f \in \{BB, EB, TB\}$ denotes a CMB field combination, $s \in \{$95$\times$95$, $95$\times$150$, $150$\times$150$\}$ denotes a spectral combination, and $j \in \{\rm{LR, GROUND, TIME, VIS}\}$ denotes a jackknife.

We test the consistency of the jackknife spectra with the null hypothesis as follows.  We estimate the diagonal covariance of the jackknife bandpowers, $\sigma^2\left(C^{fsj}_b \right)$, using the variance among the cross-spectra divided by the number of unique cross-spectra.  Next, we define jackknife ``$\chi$ bandpowers'',

\be
\chi^{ fsj }_b \equiv \frac{ C^{fsj}_b}{\sigma \left(C^{fsj}_b \right)}
\ee
and calculate four test statistics to determine the compatibility of this set of spectra with the null hypothesis.  These test statistics are:

\begin{itemize}
\item{ $\max_{fsj} \left( | \sum_b \chi^{ fsj }_b | \right) $ - This tests for spectra which are preferentially positive or negative across the full multipole range.}
\item{ $\max_{fsj} \left( \sum_b (\chi^{ fsj }_b)^2 \right) $ - This tests for spectra which preferentially have outlying bandpowers.}
\item{ $\max_{bfsj} \left( (\chi^{ fsj }_b)^2 \right) $ - This tests for any particularly strong outlying bandpower.}
\item{ $\sum_{bfsj} (\chi^{ fsj }_b)^2 $ - This tests for a general tendency to have outlying bandpowers.}
\end{itemize}

We compare the data values of these test statistics to those obtained using a set of 10000 zero-mean, unit-width Gaussian realizations of each $\chi^{ fsj }_b$.  
We calculate the probability to exceed (PTE) the value of each data test statistic given the values in the set of simulated test statistics.  
Finally, we define a global test statistic, $P_{\mathrm{joint}}$, the probability to simultaneously exceed all of the test statistics, and calculate the probability to exceed $1-P_{\mathrm{joint}}$, again using simulations.

As the focus of this work is the $BB$ spectrum, our nominal set of jackknife CMB field combinations is simply $\{BB\}$.  
We find that the PTEs for the four test statistics are 
0.29, 0.60, 0.15, and 0.09, and the global PTE is 0.22.
We take this as evidence that the $BB$ spectra do not have significant contamination under these jackknife tests.

We have also repeated these jackknife tests using an expanded set of CMB field combinations, $\{BB, EB, TB\}$.  Due to the high signal-to-noise imaging of $T$ and $E$ modes, these tests are in principle more difficult to pass.  In this case, the PTEs for the four test statistics are 
0.41, 0.59, 0.39, 0.02, and the global PTE is 0.14.
We take this as further evidence that the $B$-mode data used in this work does not have significant contamination under these jackknife tests.

\subsection{Systematic Uncertainties}
\label{sec:systematics}

Here we discuss several potential sources of systematic uncertainty in the $BB$ power spectrum measurement.  
In all cases we demonstrate
that the systematic uncertainty is much smaller than the statistical uncertainties and can be safely ignored.

\subsubsection{Uncertainty in Additive $BB$ Bias}
As described in Section~\ref{sec:additive}, we use simulations to determine the additive biases which must be subtracted from the measured $BB$ bandpowers.  Here we assess the accuracy with which we have determined these biases.

First we consider the uncertainty due to using a finite number of simulation realizations.  We have used 200 realizations, leading to uncertainties on the mean biases of 0.07$\sigma$ per $\ell$-bin.

Next we specifically consider \TtoB\ leakage, which is dominated by crosstalk leakage.  While the raw bias is as high as 1.0$\sigma$ in the lowest $\ell$-bin of 150$\times$150, convolutional cleaning reduces this bias to at most 0.1$\sigma$ per $\ell$-bin.  The uncertainty in the absolute temperature calibration is less than 3\% in power, resulting in a negligible 0.003$\sigma$ uncertainty in the additive bias.  Similarly, the underlying \LCDM\ $TT$ spectrum is constrained to approximately the same level of precision, and its uncertainty can be safely ignored.

Next we consider \EtoB\ leakage, which is caused primarily by the field geometry and the filtering of time-ordered data.  This leakage results in a bias as high as 0.6$\sigma$.  The uncertainty in the absolute polarization calibration is less than 5\% in power, resulting in a negligible 0.03$\sigma$ uncertainty in the additive bias.  Again, the underlying \LCDM\ $EE$ spectrum is constrained to a similar level of precision.

Finally we consider the negative noise bias from convolutional cleaning.  This bias is approximately -0.4$\sigma$ per $\ell$-bin in the $95 \times 95$ or $150 \times 150$ spectra.  However, because we use the data to generate the noise realizations used in the simulations, there is essentially zero systematic uncertainty in the level of this bias.

\subsubsection{Uncertainty in Multiplicative Bias in $BB$}
We discussed above how the uncertainty in the absolute polarization calibration results in a (negligible) uncertainty in the additive $BB$ bias from \EtoB.  It also results in an uncertainty on the amplitude of the measured $BB$ spectrum.  The uncertainty is 3\% and 4\% in power for the 150$\times$150 and 95$\times$150 spectra respectively, where nearly all of our sensitivity lies.  This results in a $\sim$0.1$\sigma$ global systematic uncertainty on the amplitude of the $BB$ spectrum.  We note that this small uncertainty does not affect the significance with which we detect $BB$ power, as the statistical uncertainties are noise-dominated, and the noise would suffer from the same mis-calibration.

\subsubsection{$T \rightarrow Q/U$~leakage}
\label{sec:tqu}
We see evidence for $0.65\% \pm 0.15\%$ leakage of $T$ into $Q$ and $U$ at 150 GHz.  
We measure this using $TQ$ and $TU$ spectra as described in C14.  
The source of this leakage is not known, although it could arise from our celestial calibration source, RCW38, being $\sim$0.5\% polarized.
Unlike previous SPTpol analyses which explicitly cleaned this leakage from the $Q$ and $U$ maps, this work relies on convolutional cleaning to remove this leakage.  We expect the convolutional cleaning process to reduce the associated additive $BB$ bias by at least a factor of 20, resulting in a bias that is at most $\sim$0.15$\sigma$ in the lowest $\ell$-bin of 150$\times$150 and smaller elsewhere.  The leakage is smaller yet at 95 GHz.

\subsubsection{Effect of Crosstalk on Beams}
\label{sec:xtalk_beams}
Because the electrical crosstalk between detectors happens to be preferentially negative, the array-averaged temperature beam will tend to lose solid angle.  The array-averaged polarization ($Q$/$U$) beams, on the other hand, will not lose solid angle, due to partial cancellation of crosstalk from sets of nearly randomly-oriented detectors.  This results in the array-averaged temperature and polarization beams differing slightly.  Throughout this work we have used the effective temperature beam as measured with observations of planets and AGN.  This implies that our polarized power spectra have been debiased using slightly incorrect beam functions, and that our simulations were performed using slightly incorrect beams.  The ratio of the effective temperature and polarization beam functions can be broken down into an $\ell$-independent mean offset and an $\ell$-dependent shape around that mean offset.  The $\ell$-independent offset is naturally accounted for when we calibrate our $TT$ and $EE$ spectra independently.  The $\ell$-dependent shape is not accounted for, but is well approximated by a linear tilt from +2\% to -2\% in power across the $\ell$ range of this work.  This results in negligible ($<0.1\sigma$) systematic uncertainties in the removal of additive and multiplicative $BB$ biases.

\subsubsection{Time-dependent Crosstalk}
In our baseline simulations, the crosstalk matrix $V_{ab}$ that encodes the coupling between detectors $a$ and $b$ was fixed.  In fact, we have evidence that there is some amount of time-variation in the crosstalk matrix.  Time-varying crosstalk could potentially introduce a different level of additive bias in the $BB$ spectrum than time-independent crosstalk does.  We have addressed this issue with a second set of simulations in which the crosstalk matrix $V_{ab}$ was allowed to vary, per observation, in a way that mimics the time-variation we observe.  We find that the resulting additive $BB$ bias is 
slightly smaller than
in the time-independent case.  The difference is 0.2$\sigma$ in the lowest $\ell$-bin of the 150 $\times$ 150 spectrum, and smaller than 0.04$\sigma$ in all other $\ell$-bins.

\subsubsection{Small-scale Beam Features}
We have used observations of Venus to measure the array-averaged beams on small scales ($r<3^{\prime}$).  The resulting polarization maps can be used to place upper limits on the \TtoB~leakage due to anomalous beam features on these small scales.  We estimate the spurious $BB$ power as $(C_{\pmb{\ell}}^{TB,\rm{Venus}} / C_{\pmb{\ell}}^{TT,\rm{Venus}})^2  C_\ell^{TT,\Lambda \rm{CDM}} / 20.$  We have scaled the \LCDM~$TT$ spectrum by the appropriate beam factor and have divided by 20, 
the minimum level of improvement that convolutional cleaning provides on simulated crosstalk-induced leakage (a stand-in for other types of convolutional leakage).
 We find that the resulting spurious $BB$ would introduce a net bias less than 0.02$\sigma$.

While the Venus maps provide an upper limit on the \TtoB~leakage from small-scale beam features, they do not address \EtoB~leakage.  However, given that the limits on \TtoB~are so small, and given that $EE$ is a factor of $\sim$50 smaller than $TT$ at these multipoles, we do not expect \EtoB~from small-scale beam features to be significant.

\subsubsection{Large-scale Beam Features}
We use observations of the moon to estimate that approximately 5\% of the beam solid angle is contained in the radial range $3^\prime < r < 25^\prime$.  If we consider a pessimistic scenario in which this portion of the beam, when averaged over all detectors, couples \TtoB~with 5\% efficiency or \EtoB~with 50\% efficiency (both in map units), and that convolutional cleaning reduces the resulting $BB$ power by a factor of 20, the residual $BB$ power is less than 0.03$\sigma$ for all $\ell$-bins.

\subsubsection{Systematics Summary}
To summarize, we have considered a number of systematic uncertainties and demonstrated that they are small compared to the statistical uncertainties.  When added in quadrature with the statistical uncertainties, the systematic uncertainties associated with additive biases would increase our total uncertainty by $\sim$3\% in the lowest $\ell$-bin of the 150$\times$150 spectrum, and would have a smaller effect elsewhere.  
The systematic uncertainty in the global multiplicative bias is $\sim$3.5\%.
As discussed in Section~\ref{sec:interpretation}, this is significantly smaller than the $\sim$20\% precision with which we measure any $BB$ signal.  Additionally, this multiplicative uncertainty does not affect the significance with which we detect a $BB$ signal, as it would affect the bandpowers and their uncertainties nearly equally.

\subsection{Bandpowers}
\label{sec:bandpowers}
\begin{figure*}
\begin{center}
\includegraphics[width=0.90\textwidth]{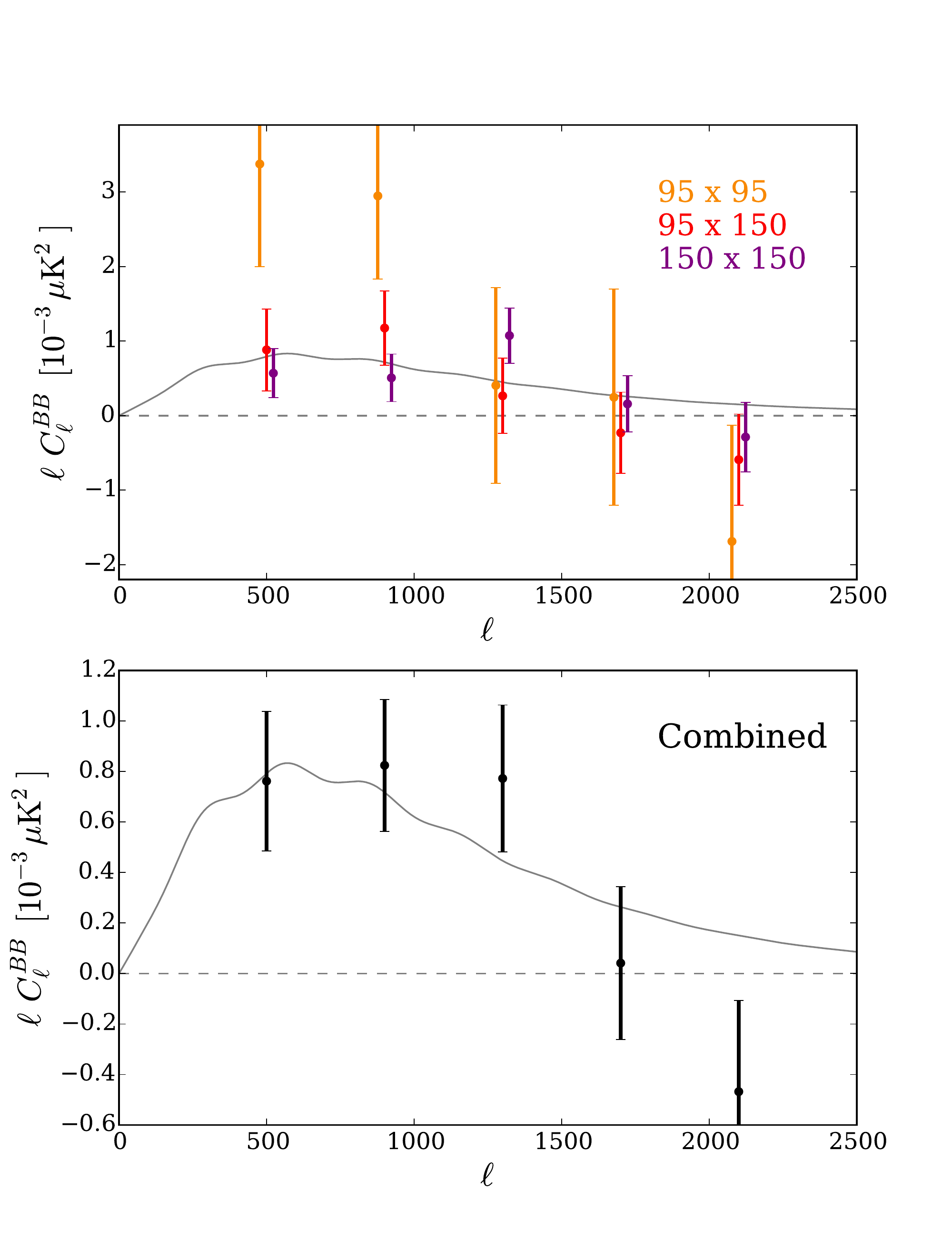}
\end{center}
\caption{
\textbf{Top}:
$BB$ power spectrum bandpowers from the individual 
95\,GHz$\times$95\,GHz,
95\,GHz$\times$150\,GHz, and
150\,GHz$\times$150\,GHz spectra.
\textbf{Bottom}:
The inverse-variance-weighted combination of the three sets of bandpowers 
in the top panel.  For reference, the expected lensed $BB$ spectrum from the \textsc{Planck+lensing+WP+highL} best-fit model in Table 5 of \cite{planck13-16} is shown by a solid gray line in each plot.
}
\label{fig:bb}
\end{figure*}

The final $BB$ bandpowers are provided in Table~\ref{table:bb} and shown in Figure~\ref{fig:bb}.  The bottom panel of Figure~\ref{fig:bb} shows the inverse-variance-weighted combination of the three sets of bandpowers.  The spectrally-combined bandpowers are for visualization purposes only; the likelihood employed in Section~\ref{sec:interpretation} uses the original set of three bandpowers.  The bandpowers, covariance matrix, and bandpower window functions are available at the SPT website\footnote{http://pole.uchicago.edu/public/data/keisler15/}.

\begin{table*}[ht!]
\caption{BB bandpowers, $\ell_{\rm{center}} C_\ell$ [$10^{-3} \mu$K$^2$]}
\label{table:bb}
\vspace{-0.2in}
\begin{center}
\begin{tabular}{ c c  | r  r |  r  r | r  r | r  r}
\hline
\hline
&&\multicolumn{2}{c}{$95 \times 95$} 
&\multicolumn{2}{c}{$95 \times 150$} 
& \multicolumn{2}{c}{$150 \times 150$}
& \multicolumn{2}{c}{Combined} \\

$\ell_{\rm{center}}$ & $\ell$ range 
& $\ell C_\ell$ & $\sigma (\ell C_\ell)$ 
& $\ell C_\ell$ & $\sigma (\ell C_\ell)$ 
& $\ell C_\ell$ & $\sigma (\ell C_\ell)$ 
& $\ell C_\ell$ & $\sigma (\ell C_\ell)$  \\
\hline
500 & 300-700 & 3.4 & 1.4 & 0.88 & 0.55 & 0.57 & 0.33 & 0.76 & 0.28 \\
900 & 700-1100 & 2.9 & 1.1 & 1.18 & 0.50 & 0.51 & 0.32 & 0.82 & 0.26 \\
1300 & 1100-1500 & 0.4 & 1.3 & 0.27 & 0.50 & 1.07 & 0.37 & 0.77 & 0.29 \\
1700 & 1500-1900 & 0.2 & 1.5 & -0.23 & 0.55 & 0.16 & 0.38 & 0.04 & 0.30 \\
2100 & 1900-2300 & -1.7 & 1.6 & -0.59 & 0.61 & -0.29 & 0.47 & -0.47 & 0.36 \\
\hline
\end{tabular}
\begin{tablenotes}
$BB$ bandpowers and uncertainties measured in this work.  The last two columns give results for the inverse-variance-weighted combination of the three sets of bandpowers.
\end{tablenotes}
\end{center}
\end{table*}

\begin{figure*}
\begin{center}
\includegraphics[width=0.98\textwidth]{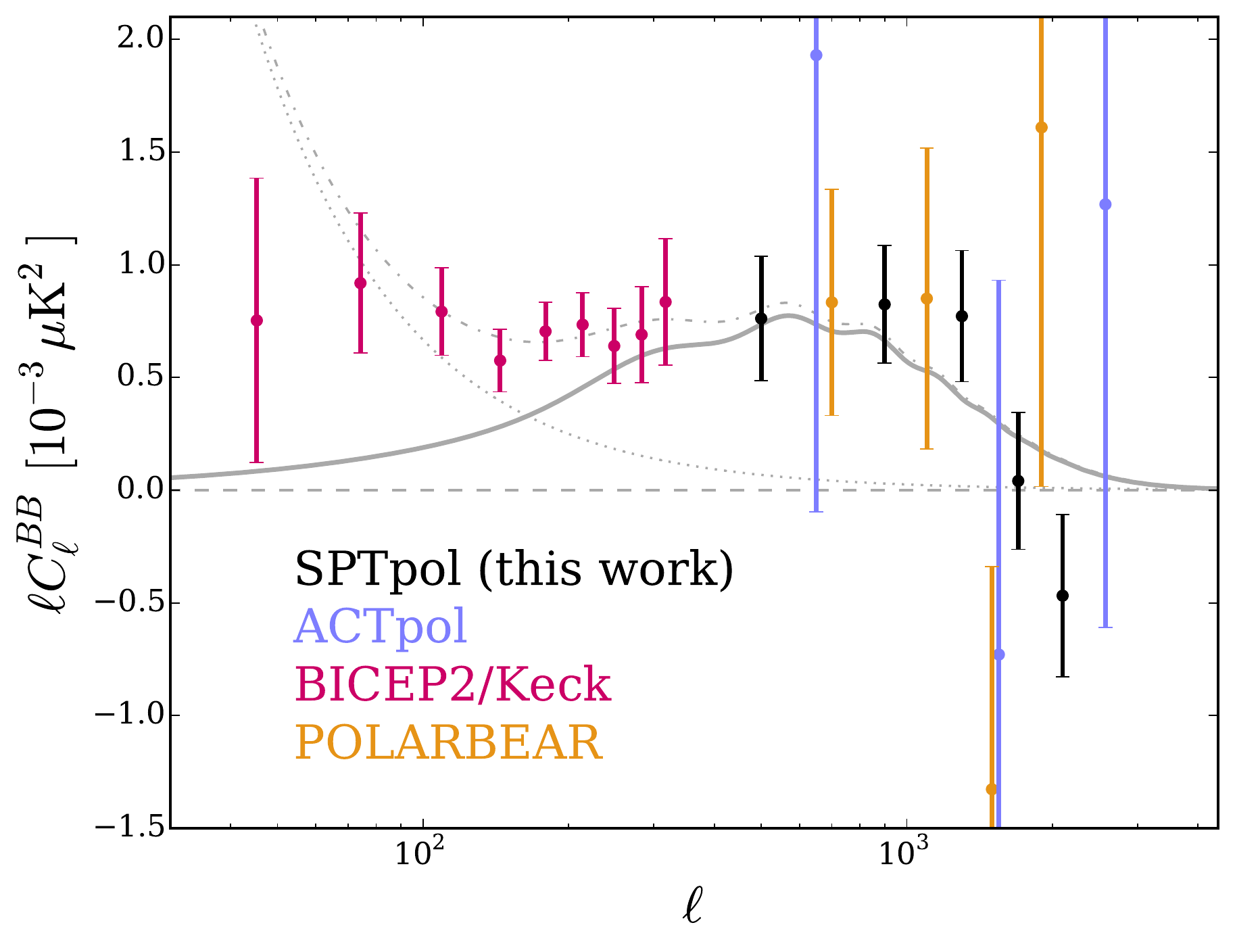}
\end{center}
\caption{
$BB$ power spectrum measurements from 
SPTpol (this work),
ACTpol \citep{naess14},
BICEP2/Keck \citep{bicep2keck15},
and
POLARBEAR \citep{polarbear2014b}.
The highest multipole bin of the ACTpol data is not shown.
The \textbf{solid gray} line shows the expected lensed $BB$ spectrum from the \textsc{Planck+lensing+WP+highL} best-fit model in Table 5 of \cite{planck13-16}.
The \textbf{dotted} line shows the 
nominal 150 GHz $BB$ power spectrum of Galactic dust emission used in this work.
This model is derived from an analysis of polarized dust emission in the BICEP2/Keck field using \planck\ data \citep{planck14-30}.
The \textbf{dash-dotted} line shows the sum of the lensed $BB$ power and dust $BB$ power.
}
\label{fig:community}
\end{figure*}

\section{Interpretation}
\label{sec:interpretation}

\subsection{Significance of BB lensing measurement}
\label{sec:bbsig}
The bandpowers from this work shown in 
Figure \ref{fig:bb} show 
a clear preference for non-zero (positive) power. 
Using a multivariate Gaussian likelihood and the bandpower covariance matrix defined in Section~\ref{sec:covariance}, we find that 
the $\chi^2$ of the data to the null hypothesis of 
zero $BB$ power is 41.5, with an associated PTE for 15 degrees of freedom (dof) of $3 \times 10^{-4}$.
The data are not well fit by zero sky power. A far better fit is the predicted lensing $BB$
spectrum from (for instance) the \textsc{Planck+Lens} best-fit model from Table II of 
\citet{planck13-16}. With zero free parameters, the $\chi^2$ of our data to this model is 18.1
for 15 dof, for a PTE of 0.25, and a $\Delta \chi^2$ relative to zero $BB$ power of 23.4.

To translate this preference into a detection significance, we introduce a single parameter, \alens, 
which we use to artificially scale the predicted lensing $BB$ power from our fiducial 
cosmological model.
We explore this parameter space (and all parameter spaces discussed subsequently in this work) using the Markov chain Monte Carlo method provided by the CosmoMC\footnote{http://cosmologist.info/cosmomc/} software package \citep{lewis02b}.
The $\chi^2$ for the best-fit model is 17.6 
($\mathrm{PTE}=0.23$).
The data prefer the addition of this single parameter by a $\Delta \chi^2$ value of 23.9, 
translating to a $4.9 \sigma$ detection of lensing, under the assumption of 
no other sky components. Note that all of these $\chi^2$ values are 
calculated with noise variance only (i.e., no sample variance from sky signals), 
because we are only asking at what level the data in 
this patch of sky prefer a component that looks like lensed $B$-modes, and are not 
attempting to relate the amplitude of this component to any global cosmological parameter.

We do expect other sky components to contribute to the measured $BB$ power, so we 
also fit for the amplitude of the lensing $B$ modes in the presence of other signals. 
We know there
will be contributions from polarized emission from extragalactic sources and from dust emission 
within our own Galaxy, and we add parameters to the fit describing each of these.

The area of sky used in this work is contained within the BICEP2 area, and should thus have a similar level of galactic polarized dust emission.  Motivated by the recent \planck\ results \citep{planck14-30}, we model the contribution from Galactic dust as:
\begin{eqnarray}
 && D^{\rm{dust}}_\ell (\nu_1 \times \nu_2) \equiv \frac{\ell (\ell+1)}{2 \pi} \ \cldust (\nu_1 \times \nu_2) = \\
\nonumber && \adust \ 
D_{80}^\mathrm{dust}(\nu_0 \times \nu_0) \,
\frac{S_\mathrm{dust}(\nu_1) S_\mathrm{dust}(\nu_2)}{S^2_\mathrm{dust}(\nu_0)}  \,
\left ( \frac{\ell}{80} \right )^{-0.42}
\end{eqnarray}
where 
$\cldust (\nu_1 \times \nu_2)$ is the contribution of dust to the 
95\,GHz$\times$95\,GHz,
95\,GHz$\times$150\,GHz, or
150\,GHz$\times$150\,GHz spectrum; 
$\adust$ is an overall scaling of the amplitude of the dust power spectrum at all
frequencies; 
$\nu_0=150\, \mathrm{GHz}$;
$D_{80}^\mathrm{dust} (\nu_0 \times \nu_0)$ is the best-fit value of the dust 
$BB$ spectrum in the BICEP2 observing region at 150\,GHz and $\ell=80$, 
according to \citet{planck14-30}; and $S_\mathrm{dust}$ is the \citet{planck14-30}
assumption for the spectral behavior of the dust (in units of CMB temperature).
We integrate the dust spectrum over the measured SPTpol bandpasses.
The only free parameter in our fit to this model is \adust.
For reference, the nominal value of \adust=1 corresponds to powers of $D_{80}^\mathrm{dust} = $  (0.00169, 0.00447, 0.0118) $\mu \rm{K}^2$ in the SPTpol spectra at (95 $\times$ 95, 95 $\times$ 150, and 150 $\times$ 150).

We model the 
contribution from extragalactic sources as a $C_\ell = \mathrm{constant}$ term, 
as would be expected if the emission was dominated by the Poisson noise in
the number of sources in our observing region, rather than the angular clustering
of those sources. We allow the amplitude of the Poisson term in our three frequency
combinations to vary independently, giving us three free parameters,
$A_\mathrm{PS,95x95}$,
$A_\mathrm{PS,95x150}$, and
$A_\mathrm{PS,150x150}$.

Based on the arguments in \citet{bicep2a} and \citet{bicep2keckplanck15}, we expect the contribution from Galactic 
synchrotron emission to be far below our ability to detect it, and we ignore this 
contribution in our analysis. We also ignore the contribution from the polarized emission
by clustered extragalactic sources. We expect this signal to be far less important relative
to the Poisson signal in polarization measurements than it is in temperature measurements,
because the clustered signal is dominated by dusty, star-forming galaxies (DSFGs), which we 
expect to have a much smaller polarization fraction than the synchrotron-emitting active
galactic nuclei (AGN)
expected to contribute the bulk of the Poisson signal \citep[e.g.,][]{seiffert07,battye11}.

Finally, we add a component to our fit that has the shape of the expected IGW $B$-mode 
signal, and we use the free parameter $r$ to control the 
the amplitude of this 
component. 
We do not expect to be able to distinguish this component from Galactic dust 
in the $\ell$ range and sensitivity level in this work; we include the IGW component
primarily to facilitate direct comparison between our lensing results and those from 
BICEP2 (see Section~\ref{sec:previous} below).

Our nominal fit includes priors on all five nuisance parameters ($A_\mathrm{PS,95x95}$,
$A_\mathrm{PS,95x150}$, $A_\mathrm{PS,150x150}$, \adust, and $r$). 
The nominal prior on \adust\ is a Gaussian centered on 1.0, with $\sigma=0.3$.
The nominal priors on Poisson power are uniform 
between zero and four times the value calculated for each parameter by
integrating source models (\citealt{dezotti10} for AGN, \citealt{negrello07} for DSFGs)
up to the unpolarized flux cut of 50\,mJy and assuming polarization fractions of $5\%$
for AGN and $2\%$ for DSFGs (the upper bounds from \citealt{seiffert07} and \citealt{battye11}, 
respectively). 
For reference, the upper edges of Poisson $C_\ell$ priors are 
$1.9\times10^{-7} \muksq$, 
$9.1\times10^{-8} \muksq$, and 
$4.4\times10^{-8} \muksq$ at 
95$\times$95, 95$\times$150, and 150$\times$150, respectively.
The prior on $r$ is uniform between 0 and 0.4. We also place a prior 
on \alens (uniform between 0 and 3.0), but the posterior on \alens\ in all fits is dominated by the 
data, not the prior.

The best-fit point in this nominal six-parameter space (including priors) has a $\chi^2$ value of
17.9. If we fix $\alens=0$ and re-fit, the best-fit point has a $\chi^2$ value of 36.8. Thus, the data
show a preference of $\Delta \chi^2 = 18.9$, or $4.3 \sigma$, for lensing $B$ modes when 
we marginalize over foreground parameters.

\subsection{Best-fit lensing amplitude and comparison with previous results}
\label{sec:previous}
To report a value of \alens\ that can be used to assess the validity of the assumed cosmological
model and to compare to other values in the literature, we repeat the six-parameter fit including
$BB$ sample variance in the uncertainty budget. The best-fit value and $1 \sigma$ uncertainty
from this fit is 
\be
\alens = 1.08 \pm 0.26
\ee
consistent with the value of 1.0 that we would expect if the 
cosmological model we assumed were correct. The constraint on \alens\ is not strongly dependent
on the foreground priors: if we increase the upper limit of the Poisson priors to 100 times the 
nominal power, the best-fit value of \alens\ shifts by less than $1 \sigma$, and the error bar 
increases by less than $10\%$. Similar behavior is seen when the \adust\ prior is changed to a uniform
prior between 0 and 3.

Figure \ref{fig:community} demonstrates that the measurement of $BB$ power in this work is 
at least visually consistent with previous measurements in the same $\ell$ range. The most 
straightforward way to compare the results quantitatively is to use the reported constraints on
\alens. This comparison is complicated somewhat by the fact that
these values are in general reported with respect to $BB$ spectra predicted using
different values of cosmological parameters. The differences between the predicted amplitudes
are typically much smaller than the $1 \sigma$ uncertainty on \alens\ from any of the measurements, 
however, and we will ignore them in the following comparison.

First we compare to other $B$-mode power spectrum measurements.  The best-fit value of \alens\ from POLARBEAR measurements of the $B$-mode power spectrum was $1.12 \pm 0.61$ \citep[statistical only,][]{polarbear2014b}, consistent with the value we
find here.  The ACTPol collaboration does not report a value of \alens\ from their $B$-mode power spectrum \citep{naess14}.
\citet{bicep2a} reports a $5.5 \sigma$ detection of lensed $B$ modes
and a best fit value of $\alens \simeq 1.75$, roughly $2 \sigma$ above 1.0, with no marginalization over foregrounds.
We have used the publicly available BICEP2 likelihood module\footnote{http://bicepkeck.org} to repeat this analysis with a marginalization over foreground and IGW tensor power that mirrors the SPTpol analysis presented here.  
We obtain a constraint from the BICEP2 data of $\alens = 1.45 \pm 0.38$, roughly 1.2$\sigma$ above 1.0.
More recently, 
\citet{bicep2keckplanck15} 
reported a strong (7$\sigma$) detection of lensing $B$-modes at $\ell \sim 200$: $\alens =1.13 \pm 0.18$.  This constraint was obtained after marginalizing over contributions from IGW tensor power and polarized galactic dust emission, and represents the most precise direct measurement of lensing $B$-modes to date.
To summarize, these $BB$ power spectrum measurements are consistent with each other, and with the \LCDM\ prediction.  
These measurements suggest that the $BB$ spectra in these particular fields and at these observing frequencies are dominated by lensing $B$-modes, at least at $\ell \gtrsim 200$.  The SPTpol $BB$ spectrum presented here is particularly useful in that regard, in the sense that $\alens\ \ge 2$ --- a rough proxy for a scenario in which other sources of $B$-modes dominate --- is rejected at $>3$$\sigma$ independent of whether or not we marginalize over foregrounds.

Next we compare to measurements of the lensing $B$-mode power spectrum that rely on cross-correlation with tracers of the CMB lensing potential $\phi$.  In these analyses, a lensing $B$-mode template is constructed by lensing measured $E$ modes by an estimate of $\phi$ derived from CIB maps or CMB lensing.  The $B$-mode template is then correlated with the measured $B$ modes to estimate the lensing $BB$ power spectrum.  The linear relationship between the CIB and $\phi$ is, in turn, based on the $C^{\phi-\rm{CIB}}_\ell$ spectrum measured by \planck, and it would therefore be surprising if these CIB cross-correlation analyses gave results that were not consistent with the $\phi\phi$ spectrum measured by Planck (which itself is consistent with the \LCDM prediction).  The best-fit value of \alens\ from the SPTpol CIB cross-correlation analysis in H13 was $1.092 \pm 0.141$.   The POLARBEAR collaboration does not report a value of \alens\ from their CIB cross-correlation analysis.  The ACTPol collaboration measured $\alens = 1.30 \pm 0.40$ in their cross-correlation analysis with \planck\ CIB data \citep{vanengelen14b}.  
\citet{planck15-15} also detect lensed $B$ modes in this fashion and 
report best-fit values of $\alens = 0.93 \pm 0.10$ when using the CIB as a $\phi$ tracer
or $\alens = 0.93 \pm 0.08$ when using the quadratic-estimator-derived $\phi$.
These measurements are consistent with the \LCDM\ prediction and with our measured value.

Finally, we note that our constraint on \alens\ is also consistent with determinations of this parameter
from quadratic-estimator reconstructions of the lensing potential using the 
four-point function of CMB temperature and polarization data. 
The strongest such constraint \citep{planck15-15} uses full-mission \planck\ data to yield $\alens = 0.983 \pm 0.025$.
Additionally, three recent works use polarization-only quadratic estimators to constrain \alens.
\citet{polarbear2014a} find $\alens = 1.06 \pm 0.47$ (statistical only), 
\citet{story14} use SPTpol data to measure 
$\alens = 0.92 \pm 0.25$ (statistical only) on the same $100 \ \degs$ field used in this work, and
\citet{planck15-15} find $\alens = 1.252 \pm 0.350$.
Again, these measurements of \alens\ are consistent with the \LCDM\ prediction and with our measured value.

\section{Conclusion}
\label{sec:conclusion}

We have presented a measurement of the $B$-mode power spectrum ($BB$ spectrum) 
using data from 100~\degs\ of sky observed 
with SPTpol in 2012 and early 2013.
The $BB$ spectrum is estimated 
in the multipole range $300 < \ell < 2300$ for three spectral combinations: 
$95\,\mathrm{GHz} \times 95\,\mathrm{GHz}$, 
$95\,\mathrm{GHz} \times 150\,\mathrm{GHz}$, and
$150\,\mathrm{GHz} \times 150\,\mathrm{GHz}$.
These data provide the best measurement of the $B$-mode power spectrum on these angular scales to date.

Several sources
of bias---all at a level below the statistical uncertainty in the power spectrum---are identified and 
subtracted from the data. 
The resulting power spectrum is strongly inconsistent with zero power 
but consistent with 
predictions for the $BB$ spectrum arising from the gravitational lensing of $E$-mode polarization.
In a six-parameter fit that includes the predicted lensed $B$-mode spectrum 
scaled by a single parameter \alens, 
as well as contributions from Galactic dust, extragalactic sources, and any IGW $B$-mode signal, 
we find $\alens = 1.08 \pm 0.26$.
The null hypothesis of no lensed
$B$-modes is ruled out at $4.3 \sigma$ after marginalizing over foreground parameters 
($4.9 \sigma$ if foregrounds are fixed to zero).

Improved constraints on the $BB$ spectrum are expected soon from a number of ongoing CMB experiments.
For example, in December 2014 SPTpol completed the second of three years of observation of a 500 \degs\ field, 
and the resulting data will significantly improve upon the $BB$ spectrum presented in this work.
A future generation of instruments aims to further constrain inflationary $B$ modes, provide maps of lensing $B$ modes over large fractions of the sky, and constrain or measure the mass in neutrinos.  We can expect significant progress in this field in the coming decade.

\acknowledgements{
The South Pole Telescope program is supported by the National Science Foundation through grant PLR-1248097. Partial support is also provided by the NSF Physics Frontier Center grant PHY-0114422 to the Kavli Institute of Cosmological Physics at the University of Chicago, the Kavli Foundation, and the Gordon and Betty Moore Foundation through Grant GBMF\#947 to the University of Chicago.  
The McGill authors acknowledge funding from the Natural Sciences and Engineering Research Council of Canada, Canadian Institute for Advanced Research, and Canada Research Chairs program.
JWH is supported by the National Science Foundation under Award No. AST-1402161.
BB is supported by the Fermi Research Alliance, LLC under Contract No. De-AC02-07CH11359 with the U.S. Department of Energy.  
The CU Boulder group acknowledges support from NSF AST-0956135.  
This work is also supported by the U.S. Department of Energy.  
Work at Argonne National Lab is supported by UChicago Argonne, LLC, Operator of Argonne National Laboratory (Argonne). 
Argonne, a U.S. Department of Energy Office of Science Laboratory, is operated under Contract No. DE-AC02-06CH11357. 
We also acknowledge support from the Argonne Center for Nanoscale Materials.  
This research used resources of the National Energy Research Scientific Computing Center, a DOE Office of Science User Facility supported by the Office of Science of the U.S. Department of Energy under Contract No. DE-AC02-05CH11231. 
The data analysis pipeline uses the scientific python stack \citep{hunter07, jones01, vanDerWalt11} and the HDF5 file format \citep{hdf5}.
}

\bibliography{../../BIBTEX/spt}

\appendix
\label{appendix:conv_cleaning_bias}
We provide here a simplified calculation of the negative, additive $BB$ bias caused by the convolutional cleaning process.
We demonstrate that the bias is negative, with a magnitude and shape similar to what we measure using simulations.
To be clear, we correct for the bias using simulations, not the results of this calculation.
The following arguments apply for (95 GHz $\times$ 95 GHz) and (150 GHz $\times$150 GHz) spectra.  We have no evidence for and do not expect a negative noise bias at (95 GHz $\times$ 150 GHz), because the noise is uncorrelated between the two spectral bands.

Consider the uniformly-weighted $BB$ bandpower at $\ell$-bin $b$:

\be
C^{BB}_{b} = 
\frac{1}{ n_{b} n_{\rm{pairs}}}
\sum_{\pmb{\ell} \in b}
\sum_{i,j \ne i}
\mathrm{Re}{ \left\{ (B^{i}_{\pmb{\ell}})^{*} B^{j}_{\pmb{\ell}} \right\} }
\ee
where $\pmb{\ell}$ is the two-dimensional Fourier space vector, $n_{b}$ is the number of two-dimensional Fourier grid points that belong in $\ell$-bin $b$, $i$ and $j$ are map bundle indices, and $n_{\rm{pairs}}$ is the number of map bundle pairs.
Note that, for purposes of clarity, the notation in the main text does not include the factor of $1/n_{b}$, while we must explicitly include that factor here.

We consider an artificial example in which the $B$ maps contain only noise.
In this limit, the bundle pairs are uncorrelated and the expectation value of the cross-spectrum is zero:
\be
\langle{ C^{BB}_{b} \rangle} \propto 
\langle{ \sum_{i,j \ne i}\mathrm{Re}{ \left\{ (B^{i}_{\pmb{\ell}})^{*} B^{j}_{\pmb{\ell}} \right\} } \rangle} = 0.
\ee
We will now demonstrate that, after applying the convolutional cleaning process, the expectation of this cross-spectrum is negative, with a magnitude similar to the bias we measure using simulations.  

The cleaning process for one convolution kernel $f_{\pmb{\ell}}$ in one $\ell$-bin $b$ essentially amounts to 1) constructing $B^{\mathrm{leak}}$, the projection of the coadded $B_{\pmb{\ell}}$ onto a unit-length mode 
$\hat{g}_{\pmb{\ell}} \equiv 
\frac{ X_{\pmb{\ell}} f_{\pmb{\ell}} }
       { \sqrt{\sum_{\pmb{\ell^\prime} \in b} |X_{\pmb{\ell^\prime}} f_{\pmb{\ell^\prime}}|^2 }}
$, 
for $X\in\{T,E\}$, 
\be
B^{\mathrm{leak}}_{\pmb{\ell}} = 
\left( \sum_{{\pmb{\ell^\prime}} \in b} \mathrm{Re}{ \left\{(B^{\mathrm{coadd}}_{\pmb{\ell^\prime}})^{*} \hat{g}_{\pmb{\ell^\prime}}  \right\}} \right) 
\hat{g}_{\pmb{\ell}},
\ee
and 2) subtracting $B^{\mathrm{leak}}_{\pmb{\ell}}$ from each bundle $B^{i}_{\pmb{\ell}}$
\be
B^{i,\rm{clean}}_{\pmb{\ell}} = B^{i}_{\pmb{\ell}} - B^{\mathrm{leak}}_{\pmb{\ell}}.
\ee
The cleaned $BB$ cross-spectrum is then
\begin{align}
\begin{split}
C^{BB,\rm{clean}}_{b} &= 
\frac{1}{ n_{b} n_{\rm{pairs}}}
\sum_{\pmb{\ell} \in b}
\sum_{i,j \ne i}
\mathrm{Re}{ \left\{ 
(B^{i}_{\pmb{\ell}} - B^{\mathrm{leak}}_{\pmb{\ell}})^{*}
(B^{j}_{\pmb{\ell}} - B^{\mathrm{leak}}_{\pmb{\ell}})
\right\} } \\
&= 
\frac{1}{ n_{b} n_{\rm{pairs}}}
\sum_{\pmb{\ell} \in b}
\sum_{i,j \ne i}
\mathrm{Re}{ \left\{ 
(B^{i}_{\pmb{\ell}})^{*} B^{j}_{\pmb{\ell}}
+ |B^{\mathrm{leak}}_{\pmb{\ell}}|^2
- \left(   (B^{i}_{\pmb{\ell}})^{*} B^{\mathrm{leak}}_{\pmb{\ell}}
   + B^{j}_{\pmb{\ell}} (B^{\mathrm{leak}}_{\pmb{\ell}})^{*}  \right)
\right\}}.
\end{split}
\end{align}
The expectation value of the first term ($\propto (B^{i}_{\pmb{\ell}})^{*} B^{j}_{\pmb{\ell}}$) is still zero.
It is straightforward to show that the expectation values of the second and third terms are 
$N / n_{b}$ and $-2N / n_{b}$, respectively,
where 
$N$ is the noise power level in the coadded $B$ map (assumed here to be $\ell$-independent, i.e., white), and
$n_b$ is the number of two-dimensional Fourier grid points belonging to $\ell$-bin $b$.
The cleaned cross-spectrum is then
\be
C^{BB,\rm{clean}}_{b} = \frac{N}{n_{b}} - \frac{2N}{n_{b}} = - \frac{N}{n_{b}}.
\ee

The difference in the expectation values of the cleaned and original power spectra, namely $(-N/n_{b} - 0) = -N/n_{b}$, is the negative additive bias caused by projecting out a single convolution kernel $f_{\pmb{\ell}}$.
When multiple, orthogonal kernels are projected out, the bias will scale as $n_{\mathrm{kernels}}$.  
The final bias is then

\be
C^{BB,\rm{bias, clean}}_{b} = - N \frac{n_{\mathrm{kernels}}}{n_{b}}.
\ee

We compare the results of this calculation to those obtained using the more realistic, simulated bandpowers.  We assume white noise levels of 17 and 9 $\mu$K-arcmin at 95 and 150 GHz, 
$n_{\mathrm{kernels}}=18$, and  
$n_{b} \simeq f_{\mathrm{sky}}(\ell_{b,\mathrm{max}}^2 - \ell_{b,\mathrm{min}}^2)$.  
The resulting biases are somewhat smaller than 
those obtained using simulations $(\mathrm{theory}/\mathrm{sims}\sim0.6)$, but the sign, shape, and overall magnitude of the two methods are in broad agreement.
This demonstrates that the basic mechanism of the bias can be understood using this simplified calculation, despite ignoring details such as non-white noise and non-uniform Fourier weights.

\end{document}